\documentclass[12pt]{article}
\usepackage{epsfig,amsfonts,amssymb}
\usepackage{hyperref}
\pdfoutput=1

\usepackage{cite}
\usepackage{cite}

\usepackage{comment}

\def\ZZZ{{\hbox{ Z\kern-1.6mm Z}}}
\def\RRR{{\hbox{ R\kern-2.4mm R}}}
\def\CCC{{\hbox{ C\kern-2.0mm C}}}
\def\zzz{{\hbox{z\kern-1mm z}}}

\newcommand{\qeq}{{\hbox{=\kern-2.3mm ? \kern.5mm }}}
\renewcommand{\qeq}{=}

\newcommand{\bF}{{\bar F}}

\newcommand{\BB}{{\cal B}}

\newcommand{\GG}{{\cal G}}

\newcommand{\FF}{{\cal F}}

\newcommand{\QQ}{{\cal Q}}
\newcommand{\PP}{{\cal P}}

\newcommand{\LL}{{\cal L}}

\newcommand{\wt}{\widetilde}
\newcommand{\wh}{\widehat}

\newcommand{\RR}{{\cal R}}

\newcommand{\TT}{{\cal T}}
\newcommand{\bg}{\bar g}

\newcommand{\be}{\begin{equation}}
\newcommand{\ee}{\end{equation}}
\newcommand{\ben}{\begin{eqnarray}\displaystyle}
\newcommand{\een}{\end{eqnarray}}

\newcommand{\refb}[1]{(\ref{#1})}
\newcommand{\p}{\partial}
\newcommand{\sectiono}[1]{\section{#1}\setcounter{equation}{0}}

\newcommand{\lsim}{\stackrel{<}{\sim}}

\def\one{{\hbox{ 1\kern-.8mm l}}}
\def\zero{{\hbox{ 0\kern-1.5mm 0}}}

\newcommand{\bea}[1]{\begin{eqnarray}\label{#1} }
\newcommand{\eea}{\end{eqnarray}}

\newcommand{\eqref}{\refb}

\usepackage{bm}
\usepackage[table]{xcolor}

\newcommand{\bR}{{\bf R}}

\def\figtwo{

\def\JPicScale{0.6}
\ifx\JPicScale\undefined\def\JPicScale{1}\fi
\unitlength \JPicScale mm

}

\def\figone{
\def\JPicScale{0.8}
\ifx\JPicScale\undefined\def\JPicScale{1}\fi
\unitlength \JPicScale mm
\begin{picture}(130,75)(0,0)
\linethickness{0.3mm}
\put(10,20){\line(1,0){120}}
\linethickness{0.3mm}
\put(10,67){\line(1,0){120}}
\linethickness{0.7mm}
\put(20,20){\line(0,1){45}}
\linethickness{0.7mm}
\put(20,65){\line(1,0){100}}
\linethickness{0.7mm}
\put(120,20){\line(0,1){45}}

\linethickness{0.1mm}
\put(22,40){\line(1,0){96}}

\put(70,44){\makebox(0,0)[cc]{$L$}}

\put(70,17){\makebox(0,0)[cc]{Horizon $z=\infty$}}

\put(70,70){\makebox(0,0)[cc]{Boundary $z=0$}}

\put(70,60){\makebox(0,0)[cc]{D3-brane}}

\end{picture}
}

\def\figthree{
 
 \def\JPicScale{0.6}
\ifx\JPicScale\undefined\def\JPicScale{1}\fi
\unitlength \JPicScale mm
\begin{picture}(130,70)(0,0)
\linethickness{0.3mm}
\qbezier(20,55)(25.17,62.89)(31.19,61.69)
\qbezier(31.19,61.69)(37.2,60.48)(45,50)
\qbezier(45,50)(52.83,39.61)(56.44,31.19)
\qbezier(56.44,31.19)(60.05,22.77)(60,15)
\qbezier(60,15)(60.08,7.17)(54.06,3.56)
\qbezier(54.06,3.56)(48.05,-0.05)(35,0)
\qbezier(35,0)(21.95,-0.06)(15.94,4.75)
\qbezier(15.94,4.75)(9.92,9.56)(10,20)
\qbezier(10,20)(9.97,30.39)(12.38,38.81)
\qbezier(12.38,38.81)(14.78,47.23)(20,55)
\linethickness{0.3mm}
\qbezier(75,40)(77.55,45.25)(82.97,45.25)
\qbezier(82.97,45.25)(88.38,45.25)(97.5,40)
\qbezier(97.5,40)(106.69,34.81)(106.69,30)
\qbezier(106.69,30)(106.69,25.19)(97.5,20)
\qbezier(97.5,20)(88.38,14.75)(82.97,14.75)
\qbezier(82.97,14.75)(77.55,14.75)(75,20)
\qbezier(75,20)(72.38,25.19)(72.38,30)
\qbezier(72.38,30)(72.38,34.81)(75,40)
\linethickness{0.3mm}
\qbezier(117.5,35)(118.79,37.62)(120.59,38.22)
\qbezier(120.59,38.22)(122.4,38.82)(125,37.5)
\qbezier(125,37.5)(127.62,36.21)(128.22,34.41)
\qbezier(128.22,34.41)(128.82,32.6)(127.5,30)
\qbezier(127.5,30)(126.21,27.38)(124.41,26.78)
\qbezier(124.41,26.78)(122.6,26.18)(120,27.5)
\qbezier(120,27.5)(117.38,28.79)(116.78,30.59)
\qbezier(116.78,30.59)(116.18,32.4)(117.5,35)
\end{picture}

 }
 
 \def\figfour{
 
 \def\JPicScale{0.8}
\ifx\JPicScale\undefined\def\JPicScale{1}\fi
\unitlength \JPicScale mm

}

\def\figoneold{

\def\JPicScale{0.4}
\ifx\JPicScale\undefined\def\JPicScale{1}\fi
\unitlength \JPicScale mm
\begin{picture}(120,75)(0,0)
\linethickness{0.3mm}
\qbezier(37.5,57.5)(46.53,66.71)(58.56,64.91)
\qbezier(58.56,64.91)(70.59,63.1)(87.5,50)
\qbezier(87.5,50)(104.58,36.99)(103.38,27.97)
\qbezier(103.38,27.97)(102.17,18.95)(82.5,12.5)
\qbezier(82.5,12.5)(62.97,5.91)(50.94,7.72)
\qbezier(50.94,7.72)(38.91,9.52)(32.5,20)
\qbezier(32.5,20)(25.92,30.38)(27.12,39.41)
\qbezier(27.12,39.41)(28.33,48.43)(37.5,57.5)
\put(105,60){\makebox(0,0)[cc]{A}}

\put(60,35){\makebox(0,0)[cc]{B}}

\end{picture}

}

\def\bF{{\bf F}}

\def\bY{{\bf Y}}

\def\bH{{\bf H}}

\def\bh{{\bf h}}

\def\bp{{\bf p}}

\def\bq{{\bf q}}

\def\bg{{\bf g}}

\begin{document}

\baselineskip 24pt

\begin{center}

{\Large \bf Decorating Asymptotically Flat  Space-Time with the Moduli Space of String Theory}

\end{center}

\vskip .6cm
\medskip

\vspace*{4.0ex}

\baselineskip=18pt

\centerline{\large \rm Ashoke Sen}

\vspace*{4.0ex}

\centerline{\large \it International Centre for Theoretical Sciences - TIFR 
}
\centerline{\large \it  Bengaluru - 560089, India}


\vspace*{1.0ex}
\centerline{\small E-mail:  ashoke.sen@icts.res.in}

\vspace*{5.0ex}

\centerline{\bf Abstract} \bigskip

N=2, 4 and 8 supersymmetric string theories in four  dimensional
flat space-time
have moduli space of
vacua. We argue that starting from a theory where the moduli approach a
particular moduli space point A  at infinity, 
we can construct a classical solution that contains 
an arbitrarily large space-time region where the moduli
take  values corresponding to any other moduli space point B of our choice to
any desired accuracy.  
Therefore the observables of a theory with a given 
set of asymptotic values of the moduli
will have complete information on the observables for any other asymptotic values
of the moduli. Also
it is physically impossible for any experiment,
performed over a finite time, to determine the asymptotic values of the moduli.
We point out the difference between asymptotically flat space-time and 
asymptotically
AdS space-time in this regard and discuss the possible implication of these results for
holographic duals of string theories in flat space-time.
For N=2 supersymmetric theories, A and B could correspond to compactifications on
topologically distinct Calabi-Yau manifolds related by flop or conifold transitions.

\vfill \eject

\tableofcontents

\sectiono{Introduction} \label{sintro}

Many quantum field theories or string theories have moduli fields
-- massless scalars
without potential.
While quantizing the theory we need to fix a boundary condition that specifies the
values of these scalars at infinity and the S-matrix of the theory depends on the
asymptotic values of these scalars. This raises the question: in what sense are the
theories corresponding to different asymptotic values of the moduli fields the same
theories?

In order to answer this question, we need to first fix some reasonable criteria
for when we can
declare two theories to be the same. We shall use the following criteria.
Let $A$ and $B$ correspond
to two different asymptotic values of the moduli and we also denote by $A$ and $B$ the
corresponding theories. Then we can declare $B$ to be a state 
in theory $A$ if any experiment
that we can perform in theory $B$ can also be performed in theory $A$.
Here by experiment we shall mean measuring an observable within some
specified error with (possibly large but) finite amount of resources.
Once we fix the desired accuracy for the experiment, this will determine the
required
size of the experimental set up and the duration of the experiment.
Let $L$ be the size of space-time that is needed to perform this experiment.
So if $A$ has a state  in which the moduli take value $B$  
to the desired accuracy in a region of space-time of size $L$, then we can
perform the desired experiment in $A$. If this can be done for every choice of $L$,
then 
$B$ can be declared a state
of $A$.

\begin{figure}
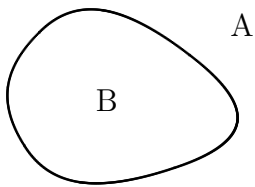

\begin{center}

\vskip -.2in

\figoneold
\end{center}

\vskip -.2in

\caption{Domain wall in a theory with asymptotic moduli $A$. \label{figoneold}}
\end{figure}

In non-gravitational
theories this can be achieved as follows.
Irrespective of what value $A$ the
moduli take at infinity, we can consider a configuration in which the
moduli take a different value $B$ of our choice in
an arbitrarily large region $\bR$ in space. 
This region $\bR$ is separated from the asymptotic region by a finite thickness
domain wall
across which the moduli change from $B$ to $A$ (see Fig.~\ref{figoneold}).
This is  an allowed 
initial configuration of the classical
fields
and hence describes a state in the theory carrying finite energy.  
The domain wall 
will evolve with time,
but if the region $\bR$ has size $L$,
then it will take a time of order $L$ for the wall to collapse. So by taking $L$ to be
sufficiently large we can ensure that we can carry out any experiment that we might
like inside the region $\bR$ to any desired accuracy and transmit the result of the
measurement to the asymptotic observer. 
Thus an observer in theory $A$ can learn
whatever they want about the theory $B$, and we can declare $B$ to be a state of $A$.

However it has been pointed out by Banks that in gravitational theories things are
more subtle\cite{2501.17697}. 
One of the issues is that a configuration where the moduli take values different
from their asymptotic values inside a large region  
may go behind an event horizon. For example in $D$ space-time
dimensions a domain wall of size $L$ will have area $L^{D-2}$ and hence mass
proportional to
$L^{D-2}$. Its Schwarzschild radius is of order $L^{(D-2)/(D-3)}$ which grows faster
than $L$ for large $L$ and hence the domain wall and the region $\bR$ will be behind the
event horizon and will be unable to communicate with the asymptotic
observer.\footnote{The triviality of the cobordism class
conjecture\cite{1909.10355} postulates that given any
two string theories A and B with same number of non-compact space-time dimensions,
there is a finite tension domain wall that separates them. The argument given above
shows that even if such a domain wall existed, 
in a theory where the asymptotic observer sees theory A, 
this domain wall cannot be used to create a large region inside which we have theory
B and which can be accessed by the asymptotic observer to carry out their
experiments. This can be circumvented if the 
domain wall tension, measured in Planck units,  
can be made parametrically small\cite{0205108}.
Note also that this argument applies only when the non-compact space-time
associated with A and B are flat
and does not preclude the possibility of a single observer having access to all string
vacua in a more general cosmological setting.}
One could get around this argument by increasing the thickness of the
wall to order $L$ so that the gradient of the scalar field $\phi$ across the wall is of order
$1/L$. In this case the domain wall tension will be of order $L(\vec \nabla\phi)^2
\sim L^{-1}$ and hence the total energy of the domain wall grows as
$L^{D-2}/L\sim L^{D-3}$. The corresponding Schwarzschild radius grows as
$L$. Since this is of the same order as the size of the system, one needs a more
careful analysis to determine whether or not the domain wall is behind the
event horizon. This will be done later.

Another argument in \cite{2501.17697} was based on holography. The asymptotic
values of the moduli fields in quantum gravity theory on AdS space-time correspond
to parameters of the dual conformal field theory and different values of these 
parameters
correspond to genuinely different conformal field theories that are not merely
different vacua of the same theory. For example, in type IIB string theory on $AdS_5\times
S^5$ the asymptotic value of the dilaton field determines the coupling constant of the
dual $N=4$ supersymmetric Yang-Mills theory. Hence 
the same must hold for the quantum gravity
theories in AdS space-time. This argument is quite robust and indeed we shall see that
our argument in flat space-time will not extend to AdS space-time.

In previous papers\cite{2502.07883,2503.00601} we showed that
using large black hole solutions in string theory in flat space-time, 
it is possible to create arbitrarily large region $\RR$ in space-time inside which the moduli
fields take values $B$ that are 
different from their asymptotic values $A$ and the space-time inside $\RR$
is locally flat, with the curvature, other field strengths and scalar field gradients
all vanishing to arbitrary accuracy. Therefore the theory $B$ 
should be regarded as a state in theory $A$. 
But a systematic study of whether all points in the moduli space can be
accessed from any given point was not addressed. This is the question we address
in this paper for $N=2,4,8$ supersymmetric theories in asymptotically flat
four dimensional space-time.  We argue that
the answer is in the affirmative, namely given any pair of points $A$ and $B$ 
in the moduli space of the theory, we can find an asymptotically flat 
classical solution where the moduli approach the point $A$ at infinity, and in
the interior we have an
arbitrarily large space-time region $\RR$  such that 
 inside $\RR$ the space-time
is flat and the moduli take the value $B$ to any desired accuracy. 
One consequence of this is that it is impossible to determine the asymptotic
values of the moduli in any experiment performed over a finite time, since we could
be living inside a space-time region where the moduli take values different from
their asymptotic values.

For definiteness 
we shall be working in theories with four non-compact space-time dimensions. However,
similar analysis can be carried out in higher dimensional theories as well. We have
chosen to work in four dimensions since the moduli spaces of these theories have
rich structure and at the same time the infrared divergences are not severe enough to
prevent us from doing such analysis. 

We also show that the
steps used to arrive at this result fail for AdS space-time. 
This is consistent with the
argument of Banks that in AdS space-time two theories that differ by the 
asymptotic values of the moduli fields should be regarded as different theories and
not different vacua of the same theory since 
this is the case in the holographic dual boundary theory. 
Reversing this logic would lead to the conclusion that
if a holographic description of string theory in flat space-time exists, then the
asymptotic values of the moduli should not appear as parameters of this dual theory.
Instead they should enter the description of the dual theory in such a way that in a
theory characterized by 
any set of values of these moduli, we should be able  to perform 
experiments to measure any property of the theory for a different set of
values of the moduli, to any desired accuracy.

Even though in string theory in AdS spaces different asymptotic values of the moduli
lead to different theories, all of these theories can be regarded as different states of
string theory in appropriate asymptotically
 flat space-time. For example if we begin with type IIA or IIB
string theory on $T^6\times R^{3,1}$, then we can construct solutions that have arbitrarily
large space-time region inside which we have type IIB on $AdS_5\times S^5$ or 
$AdS_3\times S^3\times T^4$ or M-theory
on $AdS_5\times S^7$ or $AdS_7\times S^4$. Therefore a holographic dual of type IIA
theory on $T^6\times R^{3,1}$ 
must contain complete information on these theories as well.

The rest of the paper is organized as follows. In section \ref{sgen} we describe the
strategy that we shall be using for creating arbitrarily large space-time regions inside
which the moduli take values different from their asymptotic values. In sections
\ref{sone}, \ref{stwo} and \ref{sthree} we apply this strategy respectively
to $N=4$, $N=8$ and $N=2$
supersymmetric string theories in four space-time dimensions. In section \ref{sfour} we 
analyze the special cases of topology changing transitions in type II string theories
on Calabi-Yau 3-folds and show that starting from a theory where the asymptotic
moduli correspond to compactification on a Calabi-Yau 3-fold of a particular
topology, we can create arbitrarily large regions in space-time inside which we have
compactification on a Calabi-Yau 3-fold of a different topology, related to the
original Calabi-Yau 3-fold by flop\cite{9301043,9309097} and / or 
conifold\cite{Candelas,9504090,9504145} 
transitions. Since it has been conjectured that all
Calabi-Yau 3-folds are connected by such 
transitions\cite{reid}, this result
shows that the observables of type II string theory on a particular Calabi-Yau
3-fold includes the observables of type II string theory on any other Calabi-Yau
3-fold.
Finally, in section \ref{sads} we discuss the main difference between flat space-time
and AdS space-time 
and argue that if we try to create a large region inside AdS where the
moduli take values different from their asymptotic values, this region will collapse
within a time scale set by the AdS scale. Therefore, 
from the bulk perspective, the asymptotic values
of the moduli fields in AdS should be regarded as parameters, with different values
of parameters labelling genuinely different theories. This agrees with the argument of
Banks\cite{2501.17697} from the perspective of the boundary theory.

\sectiono{General strategy} \label{sgen}

Let $A$ and $B$ be any pair of points in the moduli space.
Our goal will be to show that if the moduli approach the point $A$ at infinity, then
we can create an arbitrarily large space-time
region $\RR$ in the interior where the moduli take
values corresponding to the point $B$ and space-time curvature and all field
gradients vanish.  Furthermore we would like this
region to be outside the event horizon of any black hole that the configuration
might have, so that an experimenter can perform any desired experiment to any
desired accuracy inside the region $\RR$ 
and transmit the result to the asymptotic observer. 
This would then establish that all observables in a theory where the asymptotic
moduli take the value $B$ are also observables in the theory where the
asymptotic moduli take the value $A$, since the latter observer has access to
the observables of the former.

\subsection{Scaling}

The basic strategy for creating 
an arbitrarily large space-time region $\RR$ in the interior where the moduli take
values corresponding to the point $B$ was
outlined in \cite{2502.07883,2503.00601}. 
This is based on two steps. The first step is to identify a (possibly time dependent)
classical solution in the original theory where the moduli take the value corresponding
to the point $B$ somewhere in space-time. 
This by itself is not sufficient, since the space-time around this point may not
be locally flat and there may be other background fields ({\it e.g.} electromagnetic
fields) switched on in this region. The second step is to stretch the solution so
that the space-time becomes locally flat and all other field strengths vanish locally.
For this we use
the observation that any general covariant
action in $D$ space-time dimensions,
where each term contains two derivatives of fields, has the property that the
action is multiplied by a factor of $\lambda^{D-2}$ if we scale every rank $p$ covariant
tensor field by $\lambda^p$ and every rank $p$ contravariant tensor field
by $\lambda^{-p}$.\footnote{Such symmetries were used in \cite{9707207} 
as solution generating techniques in supergravity.}
This in particular scales the metric by $\lambda^2$ and leaves the scalars, including
the moduli, untouched. For this it is important that the scaling is done only along the
non-compact directions, so that the size parameters of the compact manifold, which are
part of the scalar moduli fields, remain unchanged under the scaling.
Also one can show that 
this transforms a $k$-brane charge $Q^{(k)}$, 
the mass $M$ and angular
momentum components $J$
as\footnote{Transformation of the $k$-brane charge may be seen as follows.
The relevant gauge field here is a $(k+1)$-form field $A^{(k+1)}$ that scales by
$\lambda^{k+1}$. The corresponding field strength $F^{(k+2)}\sim d A^{(k+1)}$ will
also scale by $\lambda^{k+1}$. Now since the scaling changes the asymptotic metric
$\eta_{\mu\nu}$ to $\lambda^2\eta_{\mu\nu}$, it is convenient to change coordinates
from $x^\mu$ to $x^{\prime\mu}=\lambda x^\mu$ so that the asymptotic metric takes the
form $\eta_{\mu\nu}dx^{\prime\mu} dx^{\prime\nu}$. 
In the new coordinates $F^{(k+2)}$ will get an
additional factor of $\lambda^{-k-2}$ and hence will scale as $\lambda^{-1}$.
We write this as $F^{\prime (k+2)}=\lambda^{-1} F^{(k+2)}$ where $F^{\prime (k+2)}$
the scaled field strength in the new coordinate system and $F^{(k+2)}$ denotes
original field strength in the original coordinate system. Now before the 
scaling, at a radial distance
$\rho$ from the brane, $F^{(k+2)}$ would have been of order $Q^{(k)}/\rho^{D-k-2}$.
Therefore $F^{\prime (k+2)} \sim \lambda^{-1} Q^{(k)}/\rho^{D-k-2}
= \lambda^{D-k-3} Q^{(k)}/\rho^{\prime(D-k-2)}$, where we used 
$\rho'=\lambda\rho$. This gives the new charge to be
$\lambda^{D-k-3} Q^{(k)}$. Similar analysis can be done for the mass and
angular momentum.
}
\be \label{echargescale}
Q^{(k)}\to \lambda^{D-k-3} Q^{(k)}, 
\qquad
 M\to \lambda^{D-3} M,
 \qquad J\to \lambda^{D-2}J\, .
 \ee
This transformation can be used to generate new solutions from
old ones. Due to the scaling of the metric, the size of the solution gets multiplied
by a factor of $\lambda$ and all invariants constructed from $k$ derivatives of various
fields scale as $\lambda^{-k}$. 
So for large $\lambda$, locally the space-time looks flat. Furthermore since the length
scale over which the scalar field varies scales as $\lambda$, by taking $\lambda$ to be
arbitrarily large we can create arbitrarily large regions where the scalar field remains
almost constant. 
Also, if the starting solution was time dependent, the time dependence 
of the new solution slows down by a factor of $\lambda$. 
This means that if the moduli take the values corresponding to
the point $B$ somewhere in space-time in the original solution, then by scaling the
solution in the manner described above, we can create an 
arbitrarily large region $\RR$
in {\it space-time} where the moduli remain almost constant at the value $B$. An
added bonus of the scaling by $\lambda$ is that
although string theory contains higher 
derivative terms, the effect of these terms become negligible for large $\lambda$.

\begin{figure}
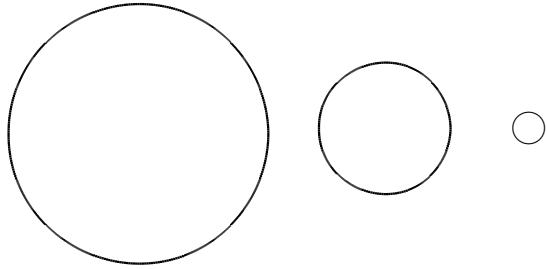

\begin{center}
\figtwo
\end{center}
\caption{Nested black holes where successive black holes are much
smaller than the previous one. However, even the smallest black hole  must
have a very large size so that the space-time around it is locally almost
flat and all fields vary slowly. \label{figtwo}}
\end{figure}

Even though the construction described above allows us to work with time
dependent solutions, in practice time dependent solutions are difficult to 
construct explicitly. We shall now describe a procedure by which we can construct
a special class of time dependent solutions on which we have full control.
Suppose that we have a time independent solution where the moduli vary as a 
function of position and let us suppose that by taking the scale parameter $\lambda$
to be large, we have created a  spatial region $\bR$ 
where the moduli take value corresponding to
a point $A'$. Now suppose that we have a second time independent
solution where the asymptotic
moduli take value corresponding to the point $A'$ and somewhere in the interior the
moduli take values $A''$. 
Suppose further that  the size of the second solution is much smaller than $\lambda$
so that from the point of view of the first solution it looks like a point. 
Then this second solution can be
put at a point inside the region $\bR$ of the first solution. 
Since the second solution is much smaller than the first solution, 
the backreaction on the first solution will be small.
At the same time, since $\lambda$ can be taken to
be arbitrarily large,  the second 
solution can also be scaled by a large parameter $\lambda'<<\lambda$. 
Even
if both the first and the second solution were time independent, the combined solution
may acquire a time dependence since the second solution
will roll in the background of the first solution over a time scale of order $\lambda$.
However by taking $\lambda$ large
we can slow down the time dependence as much as we desire.
Therefore the combined effect of this configuration will be to create an arbitrarily
large region $\RR$ in space-time inside which the moduli take values
$A''$.
This procedure can be
repeated, i.e.\ we can put a third solution in the interior of the second solution as long as
its size is small compared to $\lambda'$. This has been shown in Fig.~\ref{figtwo} using
the example of black holes.
We shall call these nested solutions. Although during much of our analysis we
shall use black holes as the basic time independent solutions for this 
construction, such nested configurations can be built from 
other objects as well, {\it e.g.} we can place a smaller loop of black string in the
background of a larger black hole.
Our goal will be to show that by considering
nested solutions of this type, we can reach any point in the moduli space from
any other point. 

One advantage of using a nested solution is that we can have analytic control on the
solution since at each stage the backreaction is small. Another advantage is that the
question of whether a given observer can send signal to the asymptotic observer, i.e.\
whether the former is outside the global event horizon, can be answered easily.
For example if we place a much smaller second black hole outside the horizon of a larger
first black hole and if we place an observer outside the horizon of the second black hole,
then this observer can certainly send signal
to the observer at infinity. 
We do not need to worry about the fact that the event horizon of the 
combined system may enclose the observer. 
This analysis can be repeated for multiple nested black holes.
On the other hand if we take a general system of multiple black holes that are
roughly of the same size, then whether an observer placed in the vicinity of this
system can send signal to infinity becomes a more complicated question. Similar
simplification occurs for other nested configurations {\it e.g.} an observer 
in a nested configuration
of string loops and black holes can send signal to infinity if at each stage the
smaller object is outside the event horizon of the larger object.

We shall end this section by describing the various kinds of configurations that we
shall be using for our analysis. 

\subsection{Black holes and strings}

Since we shall be working in four space-time dimensions,
we cannot use a membrane with spherical topology, --  its mass grows as $L^2$ if $L$
is the linear size of the system and hence its Schwarzschild radius will grow as $L^2$
which is larger than $L$. This leaves us with black holes and loops of strings. 
The latter could be either fundamental strings or various branes wrapped on
internal cycles of the compact space. 
For black
holes, by construction, as long as we are outside the event horizon we can send signal
to the asymptotic observer. We can flatten the solution by scaling the mass and
charge of the black hole by $\lambda$. We can use both extremal and non-extremal
black holes for this construction as long as we have a regular horizon with non-zero area,
since it is generally believed that existence of such black holes implies existence of 
states in the theory for which the geometry outside the horizon is described by the
black hole solution and this is the only part of the geometry that we shall be
making use of. 
But there are some BPS solutions without regular event horizon, and
the existence of the corresponding states in the theory is not
guaranteed. An example is provided by D3-branes wrapped on vanishing 3-cycles
near a conifold singularity\cite{cone,ctwo,cthree}. 
In such cases although a single wrapped D3-brane
is believed to exist\cite{9504090,9504145}, this is not so for 
multiply wrapped D3-branes and hence we 
cannot carry out the scaling \refb{echargescale} in these solutions since it
requires scaling the charges. 
This difficulty goes away
if we consider near extremal solutions since the resulting system can be regarded
as many wrapped D3-branes, gravitationally bound to a neutral black hole. 
This is important, since certain corners of the moduli space can be
reached only close to the core
of these solutions near extremality.
For practical
application it will make no difference if we replace the extremal solution by a near
extremal solution, since by tuning the extremality parameter to be sufficiently
small we can ensure that the moduli flow sufficiently close to the desired value
near the core of the solution, 
and then by scaling the solution as in \refb{echargescale} we can
ensure that the solution is as smooth as we desire.

For loops of strings\cite{Greene:1989ya,9306057} 
we have to do a more detailed analysis to determine if they
are allowed configurations since their mass grows as $L$ and hence the Schwarzschild
radius and the size of the loop scale in the same way.  We shall encounter this
issue in section \ref{sthree} where we shall use the dependence on
the string coupling $g_s$ to determine whether or not the Schwarzschild radius is smaller
than the size $L$ of the loop for small $g_s$. Similar analysis is possible in other
corners of the moduli space where tensions of some other strings 
become small. Once a configuration of this kind is constructed, we
can use the scaling transformation \refb{echargescale} to flatten the solution.
We see from \refb{echargescale} that in this case the charge of the string is
not scaled, only its size is scaled.

\subsection{Thick domain walls} \label{sthick}

Finally, we shall also make use of thick domain walls 
of the kind mentioned in section \ref{sintro}.
Let us consider a configuration where the moduli $\phi$ vary within
a sphere of radius $L$, reaching a deviation of $\Delta\phi$ from its asymptotic
value $\phi_A$ at some point $P$ inside the sphere. We shall assume that the
field configuration  is smooth, with the $n$-th spatial derivative scaling as
$\Delta\phi / L^n$. An example of such a configuration will be 
\be\label{eexample}
\phi=\phi_A + \Delta \phi \, e^{-r^2/L^2}\, ,
\ee
but our analysis below will not be restricted to this simple form.
The energy of such a configuration is
\be\label{erstwo}
{1\over 2} \int  (\vec \nabla\phi)^2 \, d^3 x \sim (\Delta\phi)^2\, L\, .
\ee
Therefore for sufficiently small $\Delta\phi$ the Schwarzschild radius of this
configuration
is smaller than the size $L$ of the configuration and an observer located within this
region will be able to communicate with the asymptotic observer.

Of course, a configuration like \refb{eexample} is not a solution to the
equations of motion and
the fields will evolve away from this initial configuration. It is easy to see using
the scaling symmetry that the time evolution will happen over a time scale $L$,
since we can get the result for a given $L$ by
starting with $L=1$ (for which the time evolution will happen over a period of order 1)
and then choosing $\lambda=L$. This changes the time evolution scale to $L$.

Let us now focus on the point $P$ where $\phi$ reaches the value $\phi_A+\Delta\phi$
and consider the limit of large $L$. In this limit we create a large region around $P$ in
which $\phi$ remains approximately at the value $\phi_A+\Delta\phi$ and this
configurations lasts for a long time. Both, the size of the region as well as the period
over which it lasts scale as $L$. So we can study the properties of the vacuum corresponding to $\phi=\phi_A+\Delta\phi$
to any desired accuracy by taking $L$ sufficiently large
and send the information back to the asymptotic observer.
Note that while the above construction works for 
$\Delta\phi$ small, the upper bound on
$\Delta\phi$ that follows from \refb{erstwo} 
is a fixed number independent of $L$. Thus taking $L$ large does not affect
how much $\Delta\phi$ we can achieve this way.

We can also consider nested configuration of such thick domain walls. 
For this we create another configuration of similar type around $P$,
with $L$ replaced by $L'<<L$. 
For this configuration $\phi_A+\Delta\phi$ will play the role of $\phi_A$,
and at some point $P'$
in the interior $\phi-\phi_A-\Delta\phi $ will reach some value $\Delta'\phi$.
$\Delta'\phi$ will also have an upper bound similar to $\Delta\phi$, but it will be
independent of $L'$. Therefore we get a net deviation of $\Delta\phi+\Delta'\phi$
from the asymptotic value of $\phi$. Since there is no upper bound on $L$ or $L'$
as long as $L'<<L$, we can repeat this process as many times as we
like to create arbitrarily large region inside which $\phi$ takes any desired value 
$\phi_B$.

We can also give a  single step procedure to make large changes in
$\phi$ as follows. Let
us take 
\be\label{e2.4rep}
\phi= \cases{ \phi_A - c\ln (r/LK) \quad \hbox{for} \quad L<r<LK, \cr
 \phi_A \quad \hbox{for} \quad r>LK, \cr
\phi_A+\Delta\phi \quad \hbox{for} \quad r<L, \quad \Delta\phi\equiv c\, \ln K\, ,
}
\ee
where $r$ is the radial variable and $L$,
$c$ and $K$
are constants.  
This gives $K=\exp[\Delta\phi/c]$. The configuration has divergent second
derivatives at $r=L$ and $LK$, but these can be smoothened out without affecting the
analysis.
Our goal is to determine the condition under which 
an observer at $r<L$ can make a measurement and send the signal to the
asymptotic observer. Suppose that the experiment is done at $r=0$ and the result is
transmitted radially outward. Then it will reach the point $r=r_0$ at a time $t\simeq r_0$.
By that time the energy contained in the $\phi$ field in the range 
$L<r<2r_0$ in the original configuration could have fallen within
the radius $r_0$. This energy is given by
 ${1\over 2}\int_L^{2 r_0} (\p_r\phi)^2 4\pi r^2 dr
=  2\pi \, c^2(2r_0-L)$ and hence its Schwarzschild radius is
$4\pi \, G\, c^2(2r_0-L)$. We need this to be less than $r_0$.  This can be 
satisfied by taking $c$ sufficiently small 
for any $r_0$ and any $\Delta\phi$, provided we take $K=\exp[\Delta\phi/c]$.
So an observer at $r<L$ can make a measurement
for values of $\phi$ that differ from the asymptotic values by $\Delta\phi$ and transmit
the signal to infinity. By taking $L$ large we can ensure that this region is sufficiently
large and lasts for sufficient amount of time so that the observer can perform the
desired experiment to any desired accuracy. Note that by taking $c$ small we not only
keep the gravitational backreaction effects small but also keep the backreaction due to all
other massless fields small since they all couple through two derivative coupling.

The construction based on thick domain wall
is very general and can be applied to any situation. 
This
gives an in principle proof that we can create any change in the moduli that we
like without encountering an event horizon.  
However  to reach points in the moduli space that
are large distance away, we either need large number of steps with
configurations of type given in \refb{eexample}, or need
exponentially large thickness
$LK=cL\exp(\Delta\phi/c)$ for configurations of type \refb{e2.4rep}.
Another somewhat unsatisfactory feature of this construction is that we do not
specify how the initial configuration is produced. This is not a serious issue, since 
any allowed initial condition for $\phi$ is a point in the phase space of the
theory and the theory
does have quantum states that approximate this, but having a known source that
can produce such configurations is useful. 
In contrast,
even though black holes and string loops do not give a universal construction that
works in all cases, they can produce large changes in $\phi$ in a single step
and they are made from known sources, -- BPS branes wrapped on various
cycles of the compact space.
For these reasons we shall use the thick domain wall sparingly
and make use of configurations of black holes and string loops whenever
possible.
A reader willing to overlook these deficiencies of the thick domain wall construction
can skip the next few sections and
jump directly to section \ref{sads}.

\sectiono{N=4 theories} \label{sone}

We begin our discussion with the N=4 supersymmetric string theories in four space-time
dimensions. The moduli spaces of these theories have simple structure, given by
$(O(6,r)/(O(6)\times O(r))) \times (SL(2,R)/U(1))$. The $O(6,r)/(O(6)\times O(r))$ moduli are
parametrized by a $(6+r)\times (6+r)$ 
symmetric $O(6,r)$ matrix $M$:\footnote{Specifying $M$
and $\tau$ fixes the point in the moduli space uniquely only after we have chosen an
underlying charge lattice. We shall assume that this has been done; {\it e.g.} for
heterotic string theory on $T^6$ for which $r=22$, we can assume that $M=I_{28}$
represents a configuration where $T^6$ is a product of six orthogonal circles, each
with radius $\sqrt{\alpha'}$, and the internal components of the Kalb-Ramond field
$B^{(2)}_{mn}$ and the $E_8\times E_8$ or $SO(32)$ gauge fields are set to zero.}
\be\label{e2.1}
M^T=M, \qquad M L M^T = L, \qquad L\equiv\pmatrix{I_6 & \cr & - I_r}\, .
\ee
The $SL(2,R)/U(1)$ moduli are parametrized by a complex number $\tau$ taking value
in the upper half plane:
\be\label{e2.2}
\tau = a + i \, e^{-2\phi}\, ,
\ee
where $\phi$ is the dilaton field and $a$ is the axion field obtained by dualizing the
Kalb-Ramond two form field $B^{(2)}_{\mu\nu}$.
There are further discrete identification of points in this moduli
space due to various duality symmetries which will not be relevant for immediate
discussion. Besides these scalar moduli, the theory also has the string frame
metric $G_{\mu\nu}$ and $6+r$ gauge fields $A_\mu^{(i)}$ for $1\le i\le 6+r$ among
its massless bosonic fields. 
The two derivative action
is invariant under the $O(6,r)$ transformation:
\be \label{e2.3}
M\to \Omega M \Omega^T, \qquad A^{(i)}_\mu \to \Omega_{ij} A_\mu^{(j)}, \qquad
G_{\mu\nu}\to G_{\mu\nu}, \qquad \tau\to \tau\, ,
\ee
where $\Omega$ is an $O(6,r)$ matrix satisfying
\be\label{e2.4}
\Omega L \Omega^T = L\, .
\ee
The equations of motion are also invariant under the $SL(2,R)$ transformation
under which
\be\label{e2.5}
\tau\to (a\tau+b)/(c\tau+d), \qquad a,b,c,d\in R, \qquad ad-bc=1\, ,
\ee
and the gauge fields transform appropriately. These transformations are in general not
genuine symmetries of the theory, -- acting on a given solution 
they general new solutions.

Our goal is to construct solutions for which the asymptotic moduli approach
some point $A$ in the moduli space, associated with specific values of $M$ and
$\tau$, and approach a point $B$ in the moduli space somewhere in the
interior of space-time. 
However by using an appropriate $O(6,r)$ transformation $\Omega$
described
in \refb{e2.3} we can bring the point $A$ to the point $A'$ corresponding to
$M=I_{6+r}$.
This will map $B$ to some point $B'$. Thus if we can 
construct a solution that approaches $M=I_{6+r}$
asymptotically and $B'$ somewhere in the interior,  then by
performing the 
$O(6,r)$ transformation $\Omega^{-1}$ on this solution, we can
 construct the desired solution.\footnote{We shall make use of black hole 
 solutions in our
construction that carry quantized charges. Therefore a general $O(6,r)\times SL(2,R)$
transformation may not take an allowed solution to an allowed solution. However since
we consider solutions with large charges, after making the $O(6,r)\times SL(2,R)$
transformation we can adjust the charges to their nearest allowed values. This does not
change the moduli significantly since they depend only on the ratios of the charges.}
Therefore we can assume, without loss of generality, that the asymptotic value of
$M$ is $I_{6+r}$.

The black hole solution that we shall use to generate the desired value of $M$
will be the electrically
charged non-rotating black hole solution. It takes the form\cite{9411187}:
\ben \label{e2.6}
ds_4^2 &=& -\Delta^{-1} \rho^2 (\rho^2-2m\rho) dt^2 + \rho^2 (\rho^2-2m\rho)^{-1} d\rho^2
+ \rho^2 d\Omega_2^2 \, , \nonumber \\
\tau &=& a_0 + i\, e^{-2\phi_0} \, \Delta^{1/2} \, \rho^{-2} \, , \nonumber \\
M &=& I_{6+r} + \pmatrix{\PP nn^T & \QQ n p^T \cr \QQ p n^T & \PP pp^T} \, , \nonumber \\
A_t &=& -{1\over \sqrt 2} \, m\, \rho\, \Delta^{-1} \, \pmatrix{
\sinh\alpha \{\cosh\beta \rho^2
+m\rho(\cosh\alpha-\cosh\beta)\} \vec n \cr
\sinh\beta \{\cosh\alpha \rho^2
+m\rho(\cosh\beta-\cosh\alpha)\} \vec p
}\, ,
\een
where 
\ben \label{e2.7}
\Delta &\equiv& \rho^4 + 2 m\rho^3(\cosh\alpha\cosh\beta-1) + m^2 \rho^2 
(\cosh\alpha-\cosh\beta)^2  \, , \nonumber \\
\PP &\equiv & 2\, \Delta^{-1} \, m^2 \rho^2 \sinh^2\alpha\sinh^2\beta\, , \nonumber \\
\QQ &\equiv & - 2\Delta^{-1}m\rho  \sinh\alpha\sinh\beta \{ \rho^2 + m\rho (\cosh\alpha\cosh\beta
-1)\}
\, .
\een
The parameters 
$m$, $\alpha$, $\beta$, the six dimensional unit vector $\vec n$ and the $r$
dimensional unit vector $\vec p$
label the mass and the $6+r$ electric charges carried by the
black hole. We take $\alpha,\beta$ to be fixed and $m$ to be large.
$a_0+i e^{-2\phi_0}$ is the asymptotic value of $\tau$ corresponding to the
moduli space point $A$.
The solution has a horizon at $\rho=2m$. So we shall take $\rho>2m$ but 
take $\rho$ to be of
the same order as $m$. 
$M$ given in \refb{e2.6} satisfies $MLM^T=L$ as a consequence
of the relation
\be
\QQ^2 = \PP(\PP+2)\, ,
\ee
which can be verified using \refb{e2.7}. Another useful relation is that $M$ given in
\refb{e2.6} can be written as
\be\label{e2.9}
M = \Omega \, \Omega^T\, ,
\ee
where
\be\label{e2.10}
\Omega = I_{6+r} + \pmatrix{Ann^T & B n p^T \cr B p n^T & App^T}\, ,
\ee
\be\label{e2.11}
B=-\Delta^{-1/2}\, m\, \rho\, \sinh\alpha\, \sinh\beta\, , 
\qquad 1+A=\Delta^{-1/2} \{ \rho^2 
+ m\rho (\cosh\alpha\cosh\beta
-1)\} \, .
\ee
Note that the scaling by $\lambda$, described in section \ref{sgen}, can be
achieved by scaling $\rho$, $m$ and $t$ by $\lambda$. $\Omega$ does not change
under this scaling, in accordance with the result that
the moduli remain unchanged under this scaling. Note also that we could
multiply the expression for $\Omega$ given in \refb{e2.10} by any
$O(6)\times O(r)$ transformation from the right without changing the 
expression for $M$ given in \refb{e2.9}. For now we proceed with \refb{e2.10} but
later we shall make use of this freedom of $O(6)\times O(r)$ multiplication from the
right.

Now  we would like to 
place another charged black hole solution (which we call the
second black hole), 
with different parameters $m',\alpha',\beta',\vec p^{\,\prime},\vec n'$ at a point
$\rho\sim m$ of the first black hole. We take $m'<<m$ so that the backreaction on
the first solution can be ignored. Let $\Omega'$ be the $O(6,r)$ matrix 
\refb{e2.10} with all the parameters replaced by the primed parameters and
$\rho$ replaced by $\rho'$. As given in \refb{e2.9}-\refb{e2.11} with primed
variables, the asymptotic value of $M$ for
the second black hole is identity and hence does not match the value 
$\Omega\Omega^T$ of $M$ at
the point $\rho$ of the first black hole. But this can be resolved by transforming the
solution associated with the second black hole by the transformation 
\refb{e2.3}, so
that the asymptotic $M$ for the second black hole becomes $\Omega\Omega^T$.
Then the value of $M$ at the point $\rho'$ of the second black hole will be
given by 
\be
\Omega \Omega'\Omega^{\prime T} \Omega^T = (\Omega \Omega')
(\Omega \Omega')^T\, .
\ee
In other words, placing a second black hole in the background of the first
black hole has the effect of composing the $O(6,r)$ transformation generated
by the two black holes. We can now repeat this analysis for a nested configuration
of black holes, reaching the conclusion that if we have $k$ nested solutions 
of type \refb{e2.6} and
if $\Omega_1,\cdots,\Omega_k$ are the $\Omega$'s given in \refb{e2.10} for these
$k$ solutions, 
then the moduli $M$ associated with the full solution will be given by
\be
M = \Omega\Omega^T, \qquad \Omega = \Omega_1\cdots \Omega_k\, .
\ee

Thus the question is whether we can get any $M$ by this construction. 
First let us count if we have enough parameters. For this note
that a general symmetric $O(6,r)$ matrix is parametrized by $6r$ parameters.
We can produce the $6r$ parameters by taking $k=6$ and 
choosing the six $\vec n_i$'s to be six
linearly independent fixed vectors, {\it e.g.} we could take $\vec n_i$ to have only the
$i$-th component non-zero. Then for each $i$, we have $r$ parameters -- one from
$B_i$ and $(r-1)$ from $\vec p_i$. This gives altogether $6r$ parameters.

We also need to show that these parameters are independent. For this we shall
consider the case where the $B_i$'s are small. In this case we have
\be
\Omega_i \simeq  \pmatrix{I_6 & B_i n_ip_i^T\cr B_i p_i n_i^T& I_r},
\qquad \Omega_1\cdots \Omega_6 
\simeq  \pmatrix{I_6 & \sum_{i=1}^6 B_i n_ip_i^T\cr \sum_{i=1}^6 
B_i p_i n_i^T& I_r}\, .
\ee
Now choosing $\vec n_1,\cdots \vec n_6$ as linearly independent fixed vectors and 
choosing the $B_i$'s and $p_i$'s appropriately, we can generate an arbitrary 
$6\times r$ matrix from $\sum_{i=1}^6 B_i n_ip_i^T$. On the other hand a general
infinitesimal symmetric $O(6,r)$ matrix has the form
\be \label{e3.15}
\pmatrix{I_6 & B\cr B^T & I_r}\, ,
\ee
where $B$ is an arbitrary $6\times r$ infinitesimal matrix. Thus we see that
by adjusting the $B_i$'s and the $p_i$'s we can generate an arbitrary infinitesimal
symmetric $O(6,r)$ matrix. This in turn shows that even for finite $B_i$, the 
parameters $\{B_i\}$ and $\{\vec p_i\}$ generically form an independent set of
$6r$ parameters
as long as the $\vec n_i$'s
are linearly independent. Therefore the matrix has the correct number of parameters
for producing a generic $M$. This does not show that any generic $M$ will be produced
this way but does show that a codimension zero subspace of
$O(6,r)/O(6)\times O(r)$ is produced this
way. 

In order to give an algorithm that can produce an arbitrary $M$, we proceed as
follows. First we use the freedom of multiplying the $\Omega_i$'s from the right
by an $O(6)\times O(r)$ transformation to multiply \refb{e3.15} by an arbitrary
$O(6)\times O(r)$ transformation close to the identity from the right. This has the
effect of adding to $I_6$ an arbitrary infinitesimal $6\times 6$ anti-symmetric matrix
and to $I_r$ an arbitrary infinitesimal $r\times r$ anti-symmetric matrix. The resulting
matrix is an arbitrary infinitesimal $O(6,r)$ matrix. Since we are not limited to
choosing infinitesimal $B_i$'s or infinitesimal $O(6)\times O(r)$ matrix, this shows that
by using the freedom of choosing arbitrary $B_i$'s and multiplying the 
$\Omega_i$'s by arbitrary $O(6)\times O(r)$
matrices from the right, we can generate any $O(6,r)$ transformation
within a ball $\BB$ of finite size
around the identity of $O(6,r)$. The desired $\Omega$  
can now be produced by taking products of elements
in $\BB$.  Such products generate only those $O(6,r)$ transformations that are
connected to the identity but this is sufficient to generate any desired $M$.
Physically the product of elements of $\BB$ can be produced by taking nested
configuration of the corresponding black holes.
The only
exceptions are the boundaries of the moduli space where (up to T-duality 
transformation) one or more of the 
radii become
large. For these, the number of required elements of $\BB$ could become large  
and one might wonder if these boundary points can be reached in finite number
of steps. This will be discussed separately later.

This construction shows how we can start with $M=I_{6+r}$ at infinity and produce
any desired value $M_B$ of $M$ somewhere in the interior. 
Next we shall discuss how we can 
get the desired value $\tau_B$ of $\tau$ in the interior. 
For this we first construct a dyonic black hole with $M=I_{6+r}$ at infinity and
carrying
electric charge vector $Q$ and magnetic charge vector $P$ such that $Q$ and $P$
only have the top 6 components non-zero and the lower $r$ components are set
to zero. In that case the solution is invariant under the transformation where we choose
$\Omega$ in \refb{e2.3} to be an $O(r)$ transformation.
As a result the moduli $M$ remain fixed at $I_{6+r}$ 
but $\tau$ varies as a function of
the radial coordinate. In particular, for an extremal supersymmetric black hole
of this type, the value of $\tau$ at the horizon takes the form\cite{9512031,0708.1270}
\be\label{e2.5a}
\tau = {Q.P\over P^2} + i \, { \sqrt{Q^2P^2 - (Q.P)^2}\over P^2}\, .
\ee
By choosing the charges $Q$ and $P$
appropriately we can get  the desired value $\tau_B$ of $\tau$ on the horizon.
We now transform this solution by an $O(6,r)$
transformation $\Omega_B$ such that $\Omega_B\Omega_B^T=M_B$. This does not
affect $\tau$ but changes $M=I_{6+r}$ to $M=M_B$ everywhere inside 
the dyonic black
hole solution. 
We can now place this dyonic black hole at a point inside the previous solution
where $M$ reaches the desired value $M_B$ starting from the asymptotic value 
$I_{6+r}$. 
As usual, to do so we need to take the dyonic black
hole to have size much smaller than the previous solution.
Near the horizon of the dyonic black hole
we now have the desired configuration $M=M_B$, $\tau=\tau_B$.

Even though we have used the near horizon geometry of an extremal dyonic
black hole to
illustrate the point, in actual practice
we do not necessarily need to use an extremal black hole or go near its
horizon. Indeed in actual practice we never work exactly on the horizon due to the
infinite red shift and time delay 
that a signal will suffer in traveling from the horizon to the
asymptotic observer.

We now discuss the points near the boundary of the moduli space. These correspond
to $\tau$ near $i\infty$ (or its images under S-duality) and / or some of the circles of
$T^6$ becoming large (or its images under T-duality). 
The question we want to address is: whether the
number of steps needed to reach the boundary points become larger as we approach
the boundary. First consider the points near $\tau=i\infty$. We see from \refb{e2.5a}
that this can be achieved in a single step by taking $\sqrt{Q^2P^2-(Q.P)^2}>> P^2$.
This in particular requires $Q^2$ to be much larger than $P^2$.

Next we turn to the configurations where one or more circles acquire large size. As
discussed in \cite{2503.00601}, this can also be achieved in a single step. 
For example, if we
want to make $d$ of the six circles to be large, we take
\ben
&& \alpha=-\beta, \qquad \vec p  = (1,c,c^2,\cdots c^{d-1}, 0,0,\cdots 0)/
\sqrt{1+c^2+\cdots c^{2d-2}}, \nonumber \\ && 
\vec n = (1,c,c^2,\cdots c^{d-1}, 0,0,\cdots 0)/
\sqrt{1+c^2+\cdots c^{2d-2}}\, ,
\een
where
\be 
 c \equiv R^{-1/d}, \qquad R\equiv (1+2m\rho^{-1} \sinh^2\alpha)^{1/2}\, ,
\ee
and take $\alpha$ to be large. This makes $R$ large and $c$ small. It was shown in
\cite{2503.00601} that in the large $R$ limit, 
all $d$ directions decompactify into circles of size
of order $R^{1/d}$. In particular any non-contractible cycle along $T^d$ has length
greater than or equal to $2\pi R^{1/d}$. 
The case when $M$ approaches other boundary points   
can be related to the one discussed
above by T-duality transformation of the solution.
This shows that approaching the points near
the boundary of the moduli space does not require us to use large number of steps.

Finally, note that  even though the construction described above gives an in principle
construction of a background producing an arbitrary value of the moduli, this is not
necessarily the minimal one. For example we could have used dyonic and / or  rotating
black holes carrying more parameters\cite{9512127} 
and hence achieved the desired goal in less
number of steps. However it is easy to see that we cannot achieve the goal with a single
black hole. Such a black hole  carries $2(r+6)$ charge parameters, 
mass and angular momentum and
there are two more parameters $(\rho,\theta)$ that label the position around the
black hole background where we work. This gives a total of $2r+16$ parameters. This
is less than the desired number of parameters $6r+2$ 
for $r> 3$. Known string theory models
have $r\ge 6$.

As already discussed earlier, the main
motivation for this construction is to show that the asymptotic observer has 
access to all the observables for all possible values of the moduli.
Indeed, once we have created an arbitrarily large region of space-time in which the
moduli take certain values, we can carry out any experiment that we desire in this
background and send the information to the asymptotic observer. 
The time taken
by such signals to travel to infinity is determined by the 
size of the largest black hole. By this time the whole set-up might collapse with
smaller black holes falling into the bigger black holes. 
However, this does not change the
fact that the asymptotic observer can gain access 
to the results of all the measurements since the region in which we perform the
experiment is outside the event horizon of the black holes.

\sectiono{N=8 theories} \label{stwo}

$N=8$ supersymmetric string theory in four space-time dimensions arises from
compactification of type IIA or IIB string theory on $T^6$. Let us for definiteness 
consider the type IIB description. In this theory 
the full moduli space is $E_{7(7)}/SU(8)$\cite{deWit,cremmer}, 
but it has an $(O(6,6)/(O(6)\times
O(6)))\times (SL(2,R)/U(1))$ subspace spanned by the vacuum expectation values of NSNS
sector fields. This has the same structure as the one for $N=4$ supersymmetric string
theories with $r=6$ and we can 
follow the procedure described in section \ref{sone} to
create large space-time
regions inside which the moduli take any value
in the $(O(6,6)/(O(6)\times
O(6)))\times (SL(2,R)/U(1))$ subspace. Therefore
we only have to discuss how to switch on RR sector moduli. These correspond to the
RR zero form field $C^{(0)}$, the components of the RR 2-form field
$C^{(2)}$ along the internal two cycles of $T^6$, the components of the RR 4-form field
$C^{(4)}$ along the internal four cycles of $T^6$ and the components of the 
RR 6-form field
$C^{(6)}$ along the whole of $T^6$. This gives altogether
\be
1 + {6\choose 2} + {6\choose 4} + 1 = 32
\ee
additional moduli. Our goal will be to understand how we could make these
moduli vary in space.

We shall start by discussing how to switch on the RR 2-form field $C^{(2)}$ along
the two cycles of $T^6$. For this we note that the moduli $M$ include the
components of the Kalb-Ramond 2-form $B^{(2)}$ along the two cycles of $T^6$. 
Since the $SL(2,Z)$ duality of type IIB in ten dimensions exchanges $B^{(2)}$ with
$C^{(2)}$, we see that 
starting with a configuration where we have switched on the moduli corresponding
to the components of $B^{(2)}$ and making an $SL(2,Z) $ transformation on this, we can
arrive at a configuration where we have switched on components of $C^{(2)}$.
By making T-duality transformation on these configurations we can construct
solutions where components of $C^{(6)}$, $C^{(4)}$ and $C^{(0)}$ are switched on.
One can now use a nested configuration of these black holes 
to switch on all the RR sector moduli besides the NSNS sector moduli.
The same arguments as in section \ref{sone} now show that
this procedure will generate the moduli
space points associated with the $E_{7(7)}$ elements within a ball $\BB$ around
the identity element. Taking products of the $E_{7(7)}$ elements inside the ball 
via nested configurations we
can generate all points in the moduli space.

\sectiono{N=2 theories} \label{sthree}

We now turn to $N=2$ supersymmetric string compactification in four space-time
dimensions. This is the minimal supersymmetry for which the theory is expected to
have a moduli space generically. 
The moduli can be divided into vector multiplets and 
hypermultiplets. We shall begin our discussion with the vector multiplets and then
turn to the hypermultiplets.

Since each vector multiplet contains 
one vector field and a complex scalar
field without potential, 
for a theory with $n_v$ vector multiplets we have altogether $n_v$ complex
or equivalently $2n_v$ real moduli fields. It is customary to label these using $n_v+1$
complex scalar fields $Y^I$ ($0\le I\le n_v$), with the physical scalars being given by the 
ratios of the $Y^I$'s. 
These are known to flow in the presence of
charged black holes. A general black hole in this theory is characterized by $n_v+1$
electric and $n_v+1$ magnetic charges, with one extra pair of charges being associated
with the graviphoton vector field of the gravity multiplet. Therefore we certainly
have enough number of parameters to make every vector multiplet moduli flow.
The issue is whether all points in the moduli space can be reached this way.

Let us denote by $\FF^I_{\mu\nu}$ the field strength associated with the $I$'th gauge
field and let us define\cite{gaillardj}
\be
(*\GG^{I})^{\mu\nu} \equiv C_0\, {\p \LL\over \p \FF^I_{\mu\nu}}\, ,
\ee
where $\LL$ denote the Lagrangian density, $C_0$ is a constant that depends
on the convention and $*$ denote the Hodge dual. 
Then the fields around
an extremal black hole solution carrying electric charges $q_I$ and magnetic
charges $p^I$  have the 
form\cite{9508072,9705169,0005049}:
\ben \label{eattract}
&& - i\, ( Y^I - \bar Y^I) = H^I(\rho), \qquad H^I(\rho)\equiv
h^I + {p^I\over \rho}\, , \nonumber \\
&& - i\,( F_I(Y) -  \bar F_I(\bar Y)) = H_I(\rho), \qquad H_I(\rho)\equiv
h_I + {q_I\over \rho},
\qquad F_I(Y)\equiv {\p F\over \p Y^I}\, , \nonumber \\
&& ds^2 = -e^{2g} dt^2 +  e^{-2g} (d\rho^2 + \rho^2 d\theta^2+\rho^2\sin^2\theta
d\phi^2),
\qquad e^{-2g}\equiv  i(\bar Y^I F_I (Y)- Y^I\, \bar F_I(Y))\, , \nonumber \\
&& \FF^I_{\theta\phi} = p^I\,\sin\theta, \qquad
 \GG^I_{\theta\phi} = q_I\sin\theta\, , \nonumber \\
&& H^I d H_I - H_I d H^I=0 \quad \Rightarrow \quad  h^I q_I- h_I p^I =  0\, ,
\een
where the prepotential $F$ is a homogeneous function of degree two in the $Y^I$'s.
$h^I$ and $h_I$ are related to the asymptotic values of $Y^I$ and $F_I$ 
using the first two equations:
\be\label{e5.2}
h^I = - i  ( Y^I_\infty - \bar Y^I_\infty), \qquad h_I = 
- i  ( F_{I\infty} - \bar F_{I\infty})\, .
\ee
Using this the last equation in \refb{eattract} can be written as
\be\label{e5.3}
Y^I_\infty  q_I- F_{I\infty}  p^I = \bar Y^I_\infty  q_I- \bar F_{I\infty}  p^I\, .
\ee
To see how we can solve these equations for given charges $q_I,p^I$ and
given asymptotic values of the vector multiplet scalar fields, 
recall that the physical scalars are
invariant under an overall complex rescaling of $Y^I$ and $F_I$. This can be
used to change the phases of the two sides of \refb{e5.3} in opposite directions
and satisfy the equality without putting any restriction on the asymptotic values
of the vector multiplet scalars. We can then use the first 
two lines in \refb{eattract}
to solve for $Y^I$'s and the third line of \refb{eattract} to find
$g$ and the metric.

The problem we want to solve is slightly different: given some arbitrary asymptotic
values of the vector multiplet scalar fields and near horizon values of the vector
multiplet scalar fields,\footnote{For definiteness we are assuming that the desired
values of the moduli are found in the near horizon region of the black hole, but
this is not necessary.}
we would like to know if there is a black hole solution
satisfying this constraint. Since as we approach the horizon $\rho=0$ the terms
involving $q_I$ and $p^I$ dominate on the right hand side of the first two equations
in \refb{eattract}, we can determine $q_I$ and $p^I$ from these equations.
However since the $Y^I$'s near the horizon are specified up to a complex 
rescaling, this determines a two parameter family of $q_I$'s and $p^I$'s, reflecting
the fact that we have $2n_v$ moduli but $2(n_v+1)$ charges.  One of these freedoms
involve the scaling of $q_I$ and $p_I$, reflecting
the scaling by $\lambda$ that was discussed in section \ref{sgen}. Once the charges
$q_I$ and $p^I$ are determined this way, we can construct the black hole solution
that has the specified values of vector multiplet moduli at infinity and near the
horizon.

One possible subtlety in the above analysis is that the path 
$Y^I(\rho)$ connecting the asymptotic
moduli $A$ and the desired moduli $B$ 
may pass through a singular point in the moduli space.
In such cases the solution may become singular at some value of
$\rho$ and may not be trustable. Such singular
subspaces of the moduli space are of real codimension two, where some period 
$m_I Y^I - n^I F_I$ vanishes for some set of
integers $\{m_I, n^I)$. As a result one can always find
a path connecting two end points avoiding these singularities. So if the path chosen by
the solution \refb{eattract} happens to pass through such singular point, we can avoid it
by considering a nested black hole configuration where the first black hole takes us from
the point $A$ to a third point $C$ and a second smaller black hole takes us from the point
$C$ to the point $B$. Another subtlety is that for certain singular points in the moduli
space the associated solution may be horizonless and possibly singular and it is not
clear if there exists states with this geometry for arbitrarily large values of the charges.
In such cases one could consider near extremal solutions by adding some neutral
matter. This will be discussed in section \ref{sfour} in the context of conifold
transitions.

This establishes that given any generic asymptotic values of the vector multiplet moduli,
we can produce any other generic values of the vector multiplet moduli near the horizon
of an appropriate black hole. Furthermore, using the freedom of scaling the
charges, we can make this region as large as we want and all invariant tensors
constructed from derivatives of the fields as small as we like. 
Although we have used extremal supersymmetric black holes 
to give a proof of this statement, 
since even for non-extremal black holes 
the moduli vary as we move along the radial
direction, in actual practice we do not need to use extremal black holes or go
all the way to the horizon, -- we just need to construct a black hole where somewhere
along the radial direction we reach the desired values of the moduli.

We now turn to the hypermultiplet moduli. These do not flow in the
black hole geometry, -- they have
the same value everywhere. We shall now discuss the mechanism for producing
large regions in the interior of space-time where the hypermultiplet
moduli take values
different from their asymptotic  values. 
We can use either type IIA or type IIB theory (on the mirror Calabi-Yau) for
describing the hypermultiplets; for definiteness let us use the type IIB
description. Let us denote by $CY$ the six real dimensional Calabi-Yau manifold on which
type IIB string theory is compactified, by $K_i$ for $1\le i\le h_{11}$ the 2-cycles
of $CY$ and by $\wt K_i$ the dual 4-cycles satisfying $K_i \cap \wt K_j=\delta_{ij}$.
Let $C^{(k)}$ denote the $k$-form RR gauge potential for any even integer $k$,
$B^{(2)}$ denote the NSNS 2-form potential and $J$ denote the Kahler form.
Then for each $i$ in the range $1\le i\le h_{11}$ we have four moduli scalar fields
from $\int_{K_i}J$, $\int_{K_i} B^{(2)}$, $\int_{K_i} C^{(2)}$ and $\int_{\wt K_i} C^{(4)}$. 
These $4\, h_{11}$ scalars
form the scalar fields of the $h_{11}$ hypermultiplet fields. Some of these can also
be given natural dual description. For example in the 3+1 dimensional 
non-compact space-time,  the gradient of the scalar
$\int_{K_i} C^{(2)}$ is the Hodge dual of the exterior derivative of the 2-form
$\int_{\wt K_i} C^{(6)}$ and the gradient of the scalar 
$\int_{\wt K_i} C^{(4)}$ is the Hodge
dual of the exterior derivative of the 2-form
$\int_{K_i} C^{(4)}$. Beside these $h_{11}$ hypermultiplets, we also have a universal
hypermultiplet whose scalars are the four dimensional dilaton $\phi_4$, the axion $a$
whose gradient is dual to the exterior derivative of the 2-form field $B^{(2)}$, the
RR 0-form potential $C^{(0)}$ and $\int_{CY} C^{(6)}$, whose gradient is dual to the
exterior derivative of the RR 2-form field $C^{(2)}$.\footnote{This description 
can be used to find the effective action of the hypermultiplet moduli fields
for large volume of the Calabi-Yau 3-fold and at weak string 
coupling; otherwise there are
$\alpha'$ and $g_s$ corrections to the effective action.}
Our goal will be to show that for
any asymptotic values of these $4(h_{11}+1)$ scalar fields, we can produce an arbitrarily
large region in space-time inside which these scalars take any other desired values.

One tool that is always available to us is the nested configuration of thick
domain walls. However we shall now describe another set of tools that involve
loops of strings obtained by wrapping various $p$-branes along the compact 
$(p-1)$ cycles of $CY$. 
As discussed in section \ref{sgen}, for string loops one needs to carefully
ensure that
the size of the loop is larger than its Schwarzschild radius so that the loop is not
enclosed by its event horizon. We shall first check under what conditions this is true.
Let $\TT$ be the tension of the string and $L$ be its
length, both measured in string units. Then its mass is given by $\TT\, L$. Its 
Schwarzschild radius $r_s$ in string units is $2\, G\, \TT\, L$ where $G$ is the Newton's
constant in string units. The condition that we need to satisfy is
$r_s< L/2\pi$ if the string is circular. Now if $g_s$ denotes the
string coupling then we have $G\sim g_s^2$.
For a fundamental string $\TT\sim 1$ and for a D$p$-brane wrapped along a
$(p-1)$-cycle of $CY$ we have 
$\TT\sim 1/g_s$. Hence we have $r_s\sim g_s^2 L$ for the fundamental string and
$r_s\sim g_s\, L$ for wrapped D-branes, and in both cases the relation $r_s<L/2\pi$ 
is satisfied for weak string coupling. On the other hand the
NS 5-branes wrapped on $\wt K_i$
have $\TT\sim g_s^{-2}$ and hence $r_s\sim L$. Therefore a more detailed study
is needed to decide if a loop of NS 5-brane on $\wt K_i$ 
is an allowed configuration, {\it e.g.} if $\wt K_i$ has small size then $\TT$ will have
a small factor and we may be able to satisfy $r_s<L$ condition.
In the following we shall assume that the string coupling is small and use only
the loops of wrapped D-branes and fundamental string. This means that
this construction will work only when both the asymptotic moduli $A$
and the desired moduli $B$  correspond to weakly coupled theory.
Given that these string loops are non-BPS configuration, it is
only expected that in the strong coupling regime the loops of fundamental strings
and D-branes may cease to be well
defined states 
and instead loops of other branes {\it e.g.} the $(p,q)$ 7-branes and
the more exotic branes
discussed in \cite{9302038,1209.6056} may be used to make the 
hypermultiplet moduli flow.

\begin{figure}
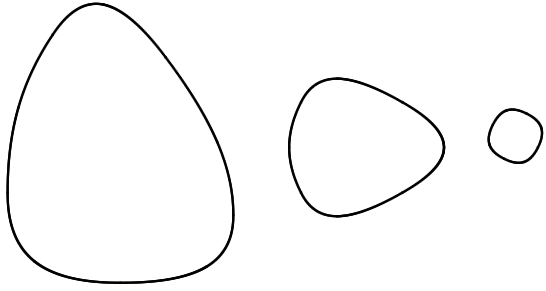

\begin{center}
\figthree
\end{center}
\caption{A `nested configuration' of  strings where each string loop is much smaller
than the previous one but all the string loops are large. According to \refb{echargescale}, 
the charge carried by each string remains fixed as we scale the size.
\label{figthree}}
\end{figure}

First we shall describe the flow of the RR sector moduli.
Consider for example
the string obtained by wrapping D3-brane on $K_i$. This is charged under the 
2-form $\int_{K_i}C^{(4)}$. Hence the dual scalar $\int_{\wt K_i} C^{(4)}$ will undergo
one unit of monodromy as we go around the string. Besides showing that 
$\int_{\wt K_i} C^{(4)}$ varies in the background of the string, this analysis also
shows that it is a periodic variable, - we shall normalize the fields so that the period
is one. Similar argument shows that $\int_{K_i} C^{(2)}$ undergoes a monodromy
as we go around a string formed by wrapping D5-brane on $\wt K_i$,
$C^{(0)}$ undergoes
monodromy around a D7-brane wrapped in $CY$
and $\int_{CY} C^{(6)}$ undergoes monodromy 
 around a D1-brane.  
 Therefore, by moving by an angle $\psi$ around a string,
 we shall produce a change
 $\Delta \phi/(2\pi)$ in the corresponding RR scalar field.
 We can now consider a nested 
 configuration of these strings as shown in \ref{figthree}, 
 with each string carrying a single (or finite) unit of charge, 
 and adjust $\psi$ for each string separately to get the desired values of the
 RR fields. Here `adjusting $\psi$' for each string means that the $(n+1)$-th string
 has to be placed at a particular point in the background of the $n$-th string and the
 experimental apparatus needs to be placed at a particular point in the background of the
 last string. 
 
We now turn to the NS sector moduli -- those corresponding to the four dimensional
dilaton $\phi_4$, the axion $a$, and the complexified Kahler moduli  
$\int_{K_i} (B^{(2)}+iJ)$. 
For this we need to study the explicit form of the string solutions. 
To simplify the analysis
we shall be  considering infinitely long straight strings, but it should be kept in
mind that in actual practice we shall be considering a large but finite
loop of string, so that close to the string it is indistinguishable from an infinitely
long string. 
The solutions can be
found using the so called C-map\cite{cmapone,cmaptwo}.\footnote{In 
type IIB on $CY$, the hypermultiplet
moduli space receives quantum corrections. The C-map described below does not
take into account these corrections. Nevertheless we expect that the qualitative picture
described below will not change, particularly if we are working at weak string
coupling.}
First let us focus on the complexified 
Kahler moduli   
$\int_{K_i} (B^{(2)}+iJ)$.
Now if, instead of IIB, we had been working with type IIA theory on the same Calabi-Yau
3-fold $CY$, then the 
complexified Kahler moduli will give the vector multiplet moduli and we would
have black hole solutions of the type given in \refb{eattract}. We 
now compactify one of the spatial directions on a circle of radius $\sqrt{\alpha'}$
and smear the charges along the circle. The resulting solution is obtained
by replacing the spherically
symmetric harmonic functions $H_I$ and $H^I$ by axially
symmetric harmonic functions. The solution takes the form
\ben \label{eattractstring}
&& - i\, ( \bY^\alpha - \bar \bY^\alpha) = \bH^\alpha(\rho), \qquad \bH^\alpha(\rho)\equiv
\bh^\alpha + {\bp^\alpha}\, \ln{\rho_0\over\rho}\, , \nonumber \\
&& - i\,( \bF_\alpha(\bY) -  \bar \bF_\alpha(\bar \bY)) = \bH_\alpha(\rho), \qquad \bH_\alpha(\rho)\equiv
\bh_\alpha + {\b\bq_\alpha}\, \ln{\rho_0\over \rho},
\qquad \bF_\alpha(\bY)\equiv {\p \bF\over \p \bY^\alpha}\, , \nonumber \\
&& ds^2 = -e^{2\bg} dt^2 + e^{-2\bg} (dz^2+d\rho^2 + \rho^2 d\psi^2),
\qquad e^{-2\bg}\equiv  i(\bar \bY^\alpha \bF_\alpha (\bY)- \bY^\alpha\, \bar \bF_\alpha(\bY))\, , \nonumber \\
&& \FF^\alpha_{z\psi} = \bp^\alpha, \qquad
 \GG^\alpha_{z\psi} = \bq_\alpha\, , \nonumber \\
&& \bh^\alpha \bq_\alpha- \bh_\alpha \bp^\alpha =  0\, ,
\een
where we have used boldfaced letters to distinguish these variables from the
complex structure moduli and gauge fields of the original type IIB theory that
appear in \refb{eattract}. $z\equiv z+2\pi\sqrt{\alpha'}$ is the coordinate 
along the circle.
\refb{eattractstring}
is not asymptotically flat since $e^{-2\bg}$ blows up as $\rho\to\infty$, but
eventually we shall consider loops of these strings for which the solution
begins to differ from \refb{eattractstring} when $\rho$ becomes of the order of the
size of the loop and the metric becomes
asymptotically flat.
The complexified Kahler moduli are given by the ratios
of the $\bY^\alpha$'s. In particular, in the large volume limit, $\bY^i/\bY^0$
for $i\ge 1$ describes the moduli $\int_{K_i} (B^{(2)}+iJ)$,
$\bp^0$ denotes the D6-brane wrapping number on $CY$, $\bp^i$ for $i\ge 1$ 
denotes
the D4-brane wrapping number on $\wt K_i$, $\bq_0$ denotes the D0-brane charge
and $\bq_i$ for $i\ge 1$ denotes the D2-brane wrapping number along $K_i$, 
with all the charges
smeared along the $z$ direction.
When the volume of the Calabi-Yau manifold is not large, 
the complexified Kahler moduli space
receives  world-sheet instanton corrections. We shall assume that the
prepotential $\bF$ appearing in \refb{eattractstring} includes these corrections.

We now make a T-duality
along this circle to convert this to a type IIB string theory on $CY\times S^1$,
with $S^1$ parametrized by the T-dual coordinate $\tilde z$.
We can now forget the periodic identification along $S^1$ and regard
the solution as describing infinitely long strings along the $\tilde z$ direction.. 
The charges $\bp^0$, $\bp^i$, $\bq_i$ and $\bq_0$ 
now become the wrapping numbers of D7-brane on $CY$, D5-branes on $\wt K_i$,
D3-brane
on $K_i$ and D1-brane, each extending along the $\tilde z$ direction.
Equivalently, $\bp^0$, $\bp^i$, $\bq_i$ and $\bq_0$ can be interpreted as the
winding numbers of $C^{(0)}$, $\int_{K_i} C^{(2)}$, $\int_{\wt K_i} C^{(4)}$ and
$\int_{CY} C^{(6)}$.
This confirms that the RR scalars
acquire monodromy around the string solution. More importantly, the first two
lines of \refb{eattractstring} show that the
complexified scalar moduli flow in the presence of the string. In particular we can
adjust their values near $\rho=0$ by adjusting the charges $\bq_\alpha$ 
and $\bp^\alpha$.\footnote{Unlike the black holes, $\rho=0$ represents a singular point
and we expect the higher derivative corrections to be important near the singularity.
However one can work at an arbitrarily small value of $\rho$ so that the scalars
are arbitrarily close to their attractor values and then scale the solution so that
the loop has large size. The scaling
in this case is achieved by taking $(t,\rho, \tilde z, \rho_0)\to
\lambda(t,\rho, \tilde z, \rho_0)$ where $\tilde z$ is the T-dual coordinate of $z$.
Of these the scaling of $(t,\rho, \tilde z)$ can be undone by coordinate change; so
the scaling of $\rho_0$ is the only relevant scaling. In this description the same values
of the complexified Kahler moduli will be reached at large values of $\rho$ where the
curvature and other field strengths are small.}
The solutions \refb{eattractstring} are corrected by string loop and D-instanton
effects since 
the hypermultiplet moduli space metric 
has such corrections, 
but as long as the string coupling $g_s$ is small, these corrections do not
change the qualitative features of the solutions.
As in the case of black holes, we do not need to go all the way to the horizon, but pick
a point where the complexified Kahler moduli take the desired values.
We can now consider a nested configuration of this string loop and the single charged
string loops described earlier to get the desired values of all the RR moduli and the
complexified Kahler moduli.

One issue that arises in this case is that since $\bp^\alpha$ and $\bq_\alpha$ are
integers while the complexified Kahler moduli space is continuous, in order to
get the correct values of the latter to a desired accuracy, we may need large $\bp^\alpha$
and $\bq_\alpha$. This was not a problem for black holes since the charges scale as
$\lambda$ and become large in the large $\lambda$ limit anyway, but for the strings
charges do not scale with $\lambda$ and taking them to be large is not natural.
In particular, if the charges become too large then the string may go behind its own
event horizon.
We can avoid this by considering a nested configuration of $2\, h_{11}$ strings, each
carrying finite charges. Each of them provides a continuous
radial parameter $\rho$ that can be
adjusted to adjust the moduli and by varying these radial parameters we can explore the
$2\, h_{11}$ dimensional complexified Kahler moduli space. In addition the angular
coordinates around these strings can be used to adjust the $2h_{11}$ RR sector
moduli. So we need an additional nested D7-brane on $CY$ and a 
D1-brane to explore the moduli space of all the RR scalars and
the complexified Kahler moduli space.

This leaves us with the modulus associated with the four dimensional dilaton $\phi_4$
and the axion $a$.
This can be achieved by using a
fundamental string. 
Near a fundamental string the axion $a$ acquires one unit of monodromy 
as we go around the string, the moduli of the Calabi-Yau manifold
remain constant and the dilaton varies as\cite{fstring}
\be
e^{-2\phi_4} \sim  \ln(\rho_0/\rho)\, .
\ee
Thus we can use the fundamental string to make $\phi_4$ flow
as we approach the core of the string. 
So by choosing an appropriate point around the fundamental string we
can achieve the desired value of $a$ and $\phi_4$.

This configuration allows us to decrease the value of $\phi_4$, i.e.\ flow
towards weak coupling, but does not allow us to increase the value of $\phi_4$. 
This is internally consistent, in that if we start from
a weakly coupled theory then we remain within the weak coupling region.
If however we want the moduli to flow towards stronger value of the coupling
then we can use
thick domain walls across which the string coupling changes from weak to strong
coupling and then use the strong-weak coupling dual of the fundamental string
loops constructed from the exotic branes mentioned earlier to make it flow towards
stronger coupling.

\sectiono{Flop and Conifold transitions} \label{sfour}

The moduli space of type IIA / IIB string theories on Calabi-Yau 3-folds 
include different components for which the topologies of the internal 
Calabi-Yau 3-folds
are different. As we move in the moduli space, we encounter subspaces
across which the topology of the internal space changes. These
are known as flop and conifold transitions. Our goal in this section will be to argue
that we can produce such transitions in physical space-time, i.e.\ by starting with 
type II string theory in which the asymptotic values of the moduli correspond
to compactification on a given internal space $CY$, we can produce a configuration
where in some arbitrarily large regions in the non-compact space-time the internal
space is a different Calabi-Yau 3-fold $CY'$ related to $CY$ by flop and / or conifold
transitions. Since it has been conjectured that all Calabi-Yau 3-folds are related by such
transitions\cite{reid}, 
this would imply that the observables of the 3+1 dimensional string theory
corresponding to some particular Calabi-Yau compactification contains information
about the observables of the 3+1 dimensional string theory associated with all other
Calabi-Yau compactifications. For definiteness we shall work with type IIB string theory
on Calabi-Yau 3-folds, -- this is of course related to type IIA string theory on the
mirror Calabi-Yau 3-folds.

The flop transition\cite{9301043,9309097} 
is easier to analyze. It appears as a topology changing transition in
the Kahler moduli space if we do not include $\alpha'$-corrections, but once we
include the world-sheet corrections the transition is smooth. This is easiest to see
in the mirror manifold where the $\alpha'$-corrected Kahler moduli space
gets mapped to the classical complex structure moduli space, and a pair of 
moduli space points on two
sides of the  flop transition can be connected by a path that does not encounter
any singularity\cite{9301043}.
Therefore the usual (possibly nested) attractor
flow in the vector multiplet moduli space via black holes and hypermultiplet moduli
space via black strings or thick domain walls 
will take us across the flop transitions.

This leaves us with the conifold transition. The physics of the transition can be
summarized in the following simple picture involving a single vector multiplet
containing a complex scalar field $\chi$ and a single hypermultiplet
whose bosonic fields involve a pair of complex scalars $H^\alpha$ ($\alpha=1,2$).
The fields have standard kinetic terms and the potential involving these fields is
of the form
\be\label{epotential}
c\, \chi^* \chi \, H^{\alpha *} H^\alpha\, .
\ee
Therefore, for $H^\alpha=0$, $\chi$ can take arbitrary vacuum expectation value. This
describes the vector multiplet moduli space, known as the deformed conifold 
branch.
On the other hand, for $\chi=0$, $H^\alpha$ can take arbitrary vacuum expectation value.
This is known as the resolved conifold or the small resolution branch.

\begin{figure}
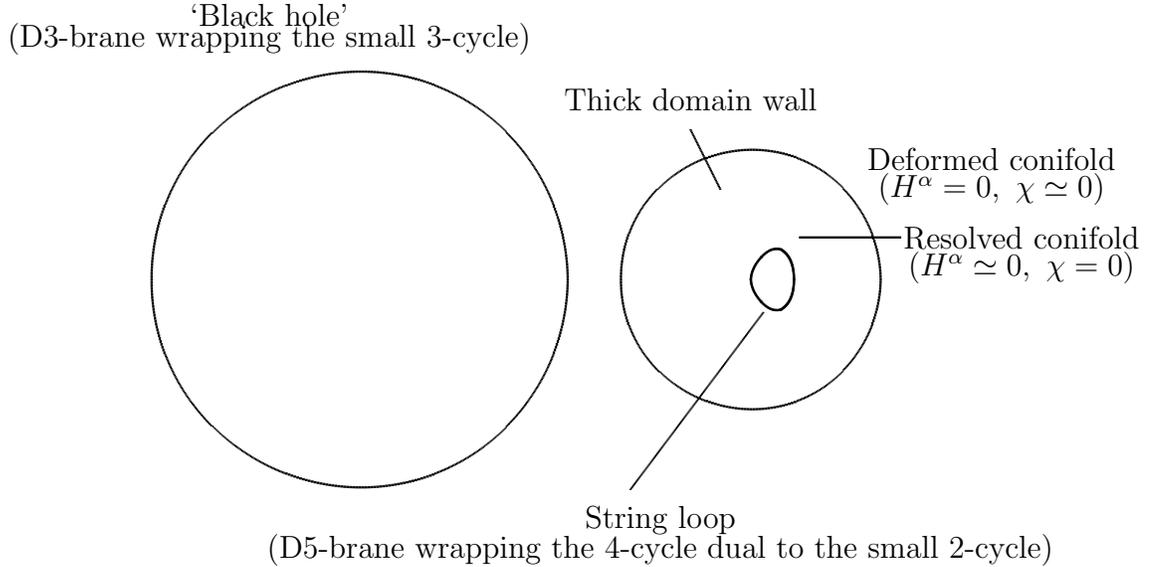

\begin{center}
\figfour
\end{center}
\caption{Nested configurations interpolating between the deformed conifold
branch and the resolved conifold branch. We have put Black hole within quotes to
emphasize that far away it looks like a black hole but near the core the
geometry differs from that of a black hole. \label{figfour}}
\end{figure}

Let us start with the deformed conifold branch where $\chi$ takes some finite value and
$H^\alpha$ vanishes. 
Our goal will be to construct a solution that has an arbitrarily large space-time region
inside which $\chi$ vanishes and $H^\alpha$ takes finite value. This will be done in three
steps, depicted in Fig.~\ref{figfour}::
\begin{enumerate}
\item We shall first construct a `black hole' solution to create a region where $\chi$ is small
and $H^\alpha$ remains zero.
\item Inside this region we shall construct a thick domain wall that interpolates between
small $\chi$, $H^\alpha=0$ configuration to $\chi=0$, small $H^\alpha$ configuration.
\item Inside the $\chi=0$, small $H^\alpha$ region we shall place a string
loop that creates
a $\chi=0$, finite $H^\alpha$ region in its vicinity.
\end{enumerate}
These configurations will be nested, in the sense that the size of each configuration
will be small compared to the previous one, but all of these configurations will
have large size.

We begin with the first step, namely construct a black hole solution where $\chi$ flows
towards zero as we approach the horizon, 
starting from an asymptotic configuration where $\chi$
is finite.
Such a solution was constructed in \cite{0005049}. It will be instructive to review this
solution. 
Let $z$ denote one of the $Y^I$'s appearing in \refb{eattract} such that 
$z=0$ corresponds to
$\chi=0$.
In this case we have, near $z=0$
\be
F_z \simeq {1\over 2\pi i} z\, \ln z\, ,
\ee
so that as $z$ goes around the origin once, $F_z\to
F_z + z$. Then \refb{eattract} gives, near $z=0$,
\ben \label{e6.3}
-i (z-\bar z) &=& h^z + {p^z\over \rho}, \nonumber \\
-{1\over 2\pi} (z \, \ln z + \bar z \, \ln \bar z) &=& h_z + {q_z\over \rho}\, , \nonumber \\
ds^2 &=& -e^{2g} dt^2 + e^{-2g} (d\rho^2 + \rho^2 d\theta^2 + \rho^2\sin^2\theta\,
d\phi^2)\, ,  \nonumber \\
e^{-2g} &=& k_1 + {1\over 2\pi} \bar z \, z\, \ln(z\bar z) \, ,
\een
where $k_1$ is a constant that represents the contribution of the other fields to
$i(\bar Y^I F_I - Y^I\bar F_I)$. We have used the gauge freedom of choosing
the overall normalization of the
$Z^I$'s such that $e^{-2g}$ approaches a finite constant $k_1$ as we
approach $z=0$.  We now choose $p^z=0$ and $h^z=0$. The former is a
choice of the charge carried by the system, while the latter implies that for
large $\rho$, $z$ becomes real. This can be ensured using the remaining gauge
freedom of multiplying all the $Z^I$'s by an overall phase. From the second 
equation in \refb{e6.3}
we now see that the $z=0$ configuration is reached at finite $\rho$:
\be
z=0 \quad \hbox{at} \quad \rho = \rho_0, \qquad \rho_0\equiv - q_z/h_z\, .
\ee
Since the $\rho=\rho_0$ sphere has non-zero size, we can continue the coordinate
$\rho$ to zero. \cite{0005049} proposed extending the solution to the $\rho<\rho_0$
region by taking $z=0$ and flat metric for $\rho<\rho_0$. The fields are continuous
but the higher derivatives of fields diverge at $\rho=\rho_0$
and so we expect the higher derivative
terms in the action to become important at $\rho=\rho_0$. For this reason we shall
not try to make use of the interior solution, but only make use of the solution for
$\rho>\rho_0$. 

In order to flatten the solution, we need to take the limit of large $q$ and $\rho$
keeping $h_z$ fixed. This makes $\rho_0$ large. Even though the solution exists,
physically we do not expect the theory to contain a bound system with arbitrarily
large charge $q$. A possible interpretation of the solution is that it represents many
copies of the $q=1$ state put together. Such a system is not bound and can split apart
into multi-centered solutions. 
We could however add a small amount of neutral matter to make the system
gravitationally bound. Irrespective of the detailed form of this solution for $\rho$ close
to $\rho_0$, one would expect that the solution in the region
$\rho>\rho_0$ will not be significantly affected by this extra matter as long 
as the extra mass
is small compared to the charge $q$. Therefore $z$ is close to 0 in this region.

Another subtlety arises from the fact that often more than one vector multiplet scalars
are involved in the conifold transition. The example described in \cite{9504145} involves
15 vector multiplet scalars $\{\chi^a\}$ and the potential \refb{epotential} is 
replaced by\footnote{There are also many hypermultiplets but only one combination has
flat potential for $\chi^a=0$. $H^\alpha$ represents this combination.}
\be\label{epotentialnew}
c\, \left( \sum_{a=1}^{15} \bar \chi^{a} \chi^a  + \sum_{a=1}^{15} \bar \chi^{a} 
\sum_{b=1}^{15} \chi^b\right)\, H^{\alpha *} H^\alpha\, .
\ee
Therefore we need to make sure that all the $\chi^a$'s become small in some region to
get a light $H^\alpha$. In this case for every $\chi^a$ we introduce a projective
coordinate field $z^a$ that form part of the $Y^I$'s and
we have $F_a = {i\over 2\pi} \left(z^a\ln z^a
+\sum_b z^b\ln \sum_c z^c\right)$ for each
$a$\cite{9504145}, and \refb{e6.3} is replaced by:
\ben \label{e6.3new}
&& -i (z^a-\bar z^a) = h^a + {p^a\over \rho}, \nonumber \\
&& -{1\over 2\pi} (z^a \, \ln z^a + \bar z^a \, \ln \bar z^a) -{1\over 2\pi}
\sum_b z^b\,  \ln \sum_c z^c
-{1\over 2\pi}
\sum_b \bar z^b \, \ln \sum_c \bar z^c
= h_a + {q_a\over \rho} \, .
\een
Now, in order to ensure that all the $z^a$'s vanish at the same value $\rho=\rho_0$,
we need to choose $p^a=0$, $h^a=0$ and $q_a/h_a=-\rho_0$ for each $a$. Unlike 
in the case of a single $z$, now the condition $h^a=0$ for each $a$ cannot be 
achieved by adjusting the overall phase of all the $Y^I$'s, -- this puts some condition on
the values of the physical moduli at large $\rho$. Similarly the condition
$q_a/h_a=\rho_0$ for each $a$ also imposes additional restriction on the moduli
fields at large $\rho$. This can be achieved using a nested configuration: given an
arbitrary set of asymptotic values of the moduli, we first use a black hole to create
a large region where the $z^a$'s are real and the imaginary parts of $F_a/q_a$
are equal for all $a$, and
then place the configuration \refb{e6.3new} in this background.

We now turn to the next step, construction of a thick domain wall.
This is a region of size $L$ outside which we have 
small $\chi$, $H^\alpha=0$ and at some point $P$ inside which we have
small $H^\alpha$, $\chi=0$. 
If inside the thick domain wall 
the fields $\chi$ and $H^\alpha$ take values of order $\phi$, then the total energy 
from the
$\int (\vec\nabla\phi)^2$ terms is of order $\phi^2 L$ and the potential energy
contribution is of order $\phi^4 L^3$. If we take $\phi\sim L^{-1}$ then both are
of order $L^{-1}$ and the backreaction of the energy density in the domain wall on
the geometry remains small for large $L$. 
We can interpret such configurations as localized
packet of $\chi$ and $H^\alpha$ waves with wavelength of order $L$, 
which is of the
same order as the inverse masses of these fields.

In the limit of large $L$ there will be
a region of size of order $L$ around the point $P$ in which $\chi$ will
remain approximately zero and $H^\alpha$ will remain 
approximately fixed at its value at $P$.
The final task is to place a string loop in this region so
that the attractor flow around the string creates a region where $H^\alpha$
goes from small value of order $L^{-1}$ to a finite value of order unity. 
(Alternatively, we could use a nested combination of thick domain walls to
achieve this.)
Physically
this corresponds to increasing the size of the 2-cycle  $K$ that is generated by
resolving the conifold since the NSNS part of $H^\alpha$ is the
complexified Kahlar modulus $\int_K (B^{(2)}+iJ)$. 
This can be done by taking the string to be a D5-brane
wrapped on the 4-cycle $\wt K$ dual to the  2-cycle $K$. To see how this works, let
us consider 
the solution \refb{eattractstring} and
choose a convention where a D5-brane wrapped on 
 $\wt K$  carries charge $\bp^1$. In that case,  $\bY^1/\bY^0$ can be identified
as $\int_K(B^{(2)}+iJ)$. 
The relevant parts of \refb{eattractstring} now take the form:
\ben \label{eattractstringnew}
&& - i\, ( \bY^1 - \bar \bY^1) =
\bh^1 + {\bp^1}\, \ln{\rho_0\over\rho}\, , \qquad 
- i\, ( \bY^\alpha - \bar \bY^\alpha) =
\bh^\alpha \quad \hbox{for $\alpha\ne 1$}\, , \nonumber \\
&& - i\,( \bF_1(\bY) -  \bar \bF_1(\bar \bY)) =0, \qquad
- i\,( \bF_\alpha(\bY) -  \bar \bF_\alpha(\bar \bY)) = \bh_\alpha  
\quad \hbox{for $\alpha\ne 1$}\,  .
\een
We now see that as
$\rho$ approaches zero, the imaginary part of $( \bY^1 - \bar \bY^1) $ grows and
hence the size of $K$ grows, achieving the desired result. Once we have
moved away from the conifold point,  
we can use further
system of black holes and string loops to reach any other desired point
in the moduli space of this new Calabi-Yau threefold.

\sectiono{Why AdS is different and implications for flat space holography} 
\label{sads}

In section \ref{sgen} we described a two step process that creates a large 
space-time region 
$\RR$ where
the moduli fields take values $B$ that are
different from their asymptotic values $A$ and the 
curvature and other field strengths vanish inside $\RR$.
The first step was to find a classical solution for which the moduli take the
values $A$ asymptotically and the
values $B$ at some interior space-time point $P$. The second step was to use the
scaling transformation  to flatten out the solution everywhere, including
the local region around the point $P$, so as to produce a large, nearly flat space-time
region inside which the moduli take the values $B$.
In AdS space-time the first step can be carried out easily. For example if we 
consider a type IIB string theory on $AdS_5\times S^5$ with $N$ units of RR five-form
flux and dilaton set to a value $\phi_0$, we can create a loop of fundamental string or
D-string in the interior of the $AdS_5\times S^5$ so that the dilaton at different points
around the loop takes different values. However the second step, that involves scaling
the solution to make the local curvature and other field strengths vanish, 
does not work. This can be seen in two ways. If we work with the effective field
theory in AdS, then the action has a cosmological constant and 
scaling transformation described in section \ref{sgen} scales the cosmological
constant term and the two derivative terms differently. Therefore it
no longer maps a classical solution to another classical solution.
On the other hand, if we work with the parent theory (ten dimensional string theory
or eleven dimensional M-theory)
then we do not have a cosmological constant to begin with, but the solution 
describing AdS has flux through compact directions that do not vanish at infinity.
The scaling transformation in the original ten or eleven
dimensional theory changes the flux and hence changes the asymptotic
boundary condition. Therefore we cannot use this to generate new solutions satisfying
the same boundary condition.

This of course does not constitute a proof that we cannot create an arbitrarily
large region
inside AdS where the moduli take values different from their asymptotic values,
but it is consistent with the observation that in the dual gauge theory the
asymptotic values of the moduli become parameters and different
asymptotic values of the moduli describe genuinely different theories\cite{2501.17697}.
Therefore, in a theory with a particular set of asymptotic values of the
moduli, we should not be able to measure the observables of a theory where the
moduli take different values asymptotically.

One can also make independent observations that illustrate the difficulty in creating
arbitrarily large domains inside $AdS$ spaces in which the moduli take values
different from their asymptotic values. 
Consider for example the possible analog of the thick domain wall solution 
described in section \ref{sthick}. 
For this we can consider the $AdS_{d+1}$ metric in global coordinates:
\be \label{eadsmetric}
ds^2 = - (1+r^2 b^{-2}) dt^2 + {dr^2 \over 1+r^2 b^{-2}} + r^2 d\Omega_{d-1}^2\, ,
\ee
where $d\Omega_{d-1}^2$ denotes the metric on a unit $(d-1)$-sphere
and $b$ is the AdS
scale. Let the thick domain wall be formed in the region $r<L$ for  some
large $L$. Now from
\refb{eadsmetric} we see that the radial ingoing light ray satisfies 
$dt=- dr/ (1+r^2 b^{-2})$, which gives
\be
t = t_0-b\, \tan^{-1} (r/b)\, .
\ee
Hence light ray can travel from $r=L$ to $r=0$ in time $b\tan^{-1}(L/b)< b\pi /2$. 
While the actual evolution of the field configuration will require a more detailed study,
this shows that the whole configuration could collapse within a period of the order of
the AdS scale instead of surviving for a time scale of order $L$.

One could also study the dynamics of the wall by regarding it as a brane and studying
its time evolution. With a later application in mind we shall study this in the Poincare
patch but a similar results holds also in the global AdS. 
In the Poincare patch we represent the metric on $AdS_{d+1}$ 
as\footnote{The global AdS
metric \refb{eadsmetric}
for large $r$ can be represented as \refb{epoinc} after identifying $z$ as $b/r$
and replacing the euclidean space spanned by the $x^i$'s by the surface of a unit
$(d-1)$-sphere. The analysis given below carries over for this case as well.}
\be\label{epoinc}
ds^2 = b^2 z^{-2} \left(dz^2 -dt^2 + \sum_{i=1}^{d-1} dx^i dx^i\right)\, .
\ee
$z=0$ represents the
boundary of $AdS_{d+1}$ space.
We represent the initial configuration of the domain wall by a
brane parallel to the boundary placed at the point $z=z_0$.
In this case the world-volume
action of the brane is proportional to:
\be
-\int d^{d}\xi\, 
\sqrt{-\det h}\, , \qquad h_{\alpha\beta} =\p_\alpha x^\mu \p_\beta x^\nu g_{\mu\nu}(x)
\, ,
\ee
where $\xi^\alpha$ for $0\le\alpha\le d-1$ are the world-volume coordinates on the
brane and $g_{\mu\nu}$ is the space-time metric. We shall choose the coordinates
$\xi^i$ for $1\le i\le d-1$ to coincide with the space-time coordinates $x^i$ so 
that $\p_i x^j=\delta_i^j$ and take $t$ and $z$ to be functions of $\tau\equiv \xi^0$.
Then the integral over the $\xi^i$'s trivially factor out and the action is
proportional to
\be
-\int d\tau\, z(\tau)^{-d} \sqrt{(\p_\tau t)^2 - (\p_\tau z)^2}\, .
\ee
Associated with the $t\to t+c$ symmetry there is a conserved Noether charge:
\be
C = z^{-d} \, {\p_\tau t\over \sqrt{(\p_\tau t)^2 - (\p_\tau z)^2}}\, .
\ee
We now choose the $\tau=t$ gauge so that we have
\be \label{ebraneeq}
C=z^{-d} \, {1\over \sqrt{(1 - (\p_t z)^2}}\, .
\ee
If we assume that the brane was at rest
at $z=z_0$ at $t=0$, then we get $C=z_0^{-d}$ and hence
\refb{ebraneeq} gives
\be
\p_t z = \sqrt{1 - (z_0/z)^{2d}}\, .
\ee
Integrating this equation we see that the brane will reach $z=z_1$ at time $t=t_1$
given by
\be\label{e7.9}
t_1 = \int_{z_0}^{z_1} { z^d dz \over  \sqrt{z^{2d}-z_0^{2d}}} = 
z_0 \int_1^{z_1/z_0} {u^d du \over \sqrt{u^{2d}-1}}\, .
\ee
For small $z_0$, \refb{e7.9} gives 
$t_1\simeq z_1$. 
\refb{epoinc} now shows that the proper time of an observer at $z_1$
will be given by $b\, t_1/z_1\simeq b$. Therefore we again see that the brane will
reach the observer at a proper time of the order of the AdS scale $b$, 
and as a result the observer
will be unable to carry out an experiment with arbitrary accuracy.

\begin{figure}
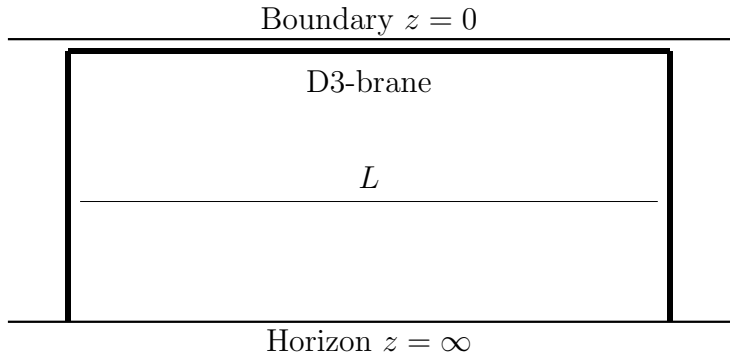

\begin{center}
\figone
\end{center}

\vskip -.6in

\caption{The effect of pulling a set of D3-branes away from the horizon
towards the boundary is to create a large region in the interior of the
Poincare patch of AdS between the 
D3-branes
and  the horizon where
the RR flux is different from its asymptotic value. The boundary dual of this is a large
region where the adjoint scalars take a large vacuum expectation value,
breaking $SU(N)$ to $SU(N-M)$\cite{9811120}. \label{figone}}
\end{figure}

There are exceptions to this, {\it e.g.} in type IIB string theory in $AdS_5\times S^5$
with RR 5-form flux turned on, one can place a finite number $M$ of D3-branes parallel
to the boundary at any point $z=z_0$.  For simplicity, we shall assume that $M$ is
large and the D3-branes are spread uniformly across the sphere $S^5$ so that we
can regard the D3-brane charge to be smeared over
$S^5$. In this case the attractive force due to gravity cancels the repulsive force
due to RR flux and the brane is static. 
This is related to the fact that in $N=4$ supersymmetric
Yang-Mills theory on $R^{3,1}$ we do have a set of moduli corresponding to the
vacuum expectation values of the adjoint scalars\cite{9811120}. 
Creating arbitrarily large regions in
the boundary where these moduli take vacuum expectation values correspond to the
ability to create arbitrarily large region in the bulk dual Poincare AdS where a set of
D3-branes are pulled away from the horizon towards the 
boundary\cite{9811120}.
This has
been shown in Fig.\ref{figone}. In this case in the bulk we have created an arbitrarily
large region where the RR flux is less than its value at infinity, and hence the gauge
group is $SU(N-M)$ for some positive $M$. This is consistent with the fact that in the
boundary theory we can create an arbitrarily large region with gauge group $SU(N-M)$
by giving large vacuum expectation values to $M$ of the components of the
adjoint scalar fields. This configuration does not live for infinitely long either in the
bulk or the boundary, but can be made to survive for arbitrarily long time by taking
the extent $L$ of the region parallel to the boundary arbitrarily large.

We can try to use these results to gain some insight into the
properties of holographic duals of
string theory in flat space-time. We have seen that 
in flat space-time we can create arbitrarily large
regions where the moduli take values different from their asymptotic values
and hence the observables in a theory corresponding to particular asymptotic
values of the moduli must contain the observables in theories corresponding to
all other asymptotic values of the moduli.
Thus the same must also hold for the holographic dual.
This shows that the asymptotic values of the moduli fields
should not appear as parameters labelling the dual theory, with different values
of the parameters labelling physically distinct theories.
Instead, within a theory corresponding to
a given set of asymptotic values of the moduli, 
we should have the means to measure the observables for 
a different choice of these asymptotic values. 

One can envisage two possible scenarios in which this can happen. One is that the
holographic description of string theory 
in flat space-time necessarily involves a large $N$ limit
as in the BFSS matrix model\cite{9610043} or limit of AdS/CFT 
correspondence\cite{9711200}. In this case
moduli remain parameters of the theory at all finite $N$ and 
the different theories corresponding to different values of these parameters
become different states of the same theory
only at infinite $N$. This will be somewhat disappointing, since
without knowing the bulk dual at finite but large $N$ it may be hard to come up with
an argument for this purely from the boundary theory. This is related to the fact
that in AdS/CFT correspondence, the emergence of bulk locality is hard to see purely
from the CFT perspective.\footnote{This
analysis also tells us that for a large $N$ CFT with a
holographic dual, labelled by a set of parameters, one should be able to device
experiments that measure properties of this CFT for different values of the
parameters in the $N\to\infty$ limit.}
The second possibility is that the asymptotic values of
the moduli fields in flat space-time actually appear as vacuum expectation values of
some moduli scalar fields even in the boundary theory. There could also be other possibilities.

We conclude this section by noting that
while we cannot apply the scaling transformation to create new
solutions in asymptotically AdS space-time, we can create AdS space-time inside
asymptotically flat space-time using a slight variation of the scaling transformation.
For example in type IIB string theory on $T^6\times R^{3,1}$, we can
produce an arbitrarily large locally flat region in the interior where the radii of the
internal circles become arbitrarily large, creating a space-time that is indistinguishable
from type IIB string theory in ten dimensional flat space-time. In this background
we can place $N$ D3-branes wrapped along three of the large internal circles. From the
point of view of the asymptotic observer these would appear as charged particles.
The main difference from all our earlier constructions is that we do not scale the D3-brane
charge to be very large. Instead we keep the number $N$ of D3-branes fixed so
that near the horizon of the D3-brane system we get locally $AdS_5\times S^5$ metric
in the Poincare patch instead of getting locally flat space-time.
Of course this configuration is not time independent since the D3-branes would roll
in the black hole background, but since the local variation of the background
fields is small, the configuration
can be long lived. 
We can then go sufficiently close to the horizon of the D3-brane system and
measure the S-matrix of type IIB string theory on $AdS_5\times S^5$ to any
desired accuracy.

By combining this type of construction with the ones described earlier
one can argue that   the observables of type IIA string theory on $T^6\times R^{3,1}$
include the observables of\cite{2503.00601}
\begin{enumerate}
\item M-theory on $R^{10,1}$, 
\item type IIA and type
IIB string
theories on $R^{9,1}$, 
\item type IIB string theory on $AdS_5\times S^5$ with any value
of the dilaton and any amount of RR 5-form 
flux, 
\item M-theory on $AdS_4\times S^7$ and $AdS_7\times
S^4$ with any amount of 4-form flux, 
\item type IIB string theory on $AdS_3\times S^3
\times T^4$ with any amount of flux, 
\end{enumerate}
and many other 
backgrounds\cite{2503.00601}. Therefore the holographic
description of type IIA string theory on $T^6\times R^{3,1}$ must also contain
information on the observables of all of these other backgrounds. 
This suggests that the holographic description of string theory in backgrounds
with flat non-compact space-time is
likely to be more challenging and more rewarding than their AdS counterpart.

\bigskip

\noindent{\bf Acknowledgement:} 
I wish to thank Tom Banks,  Tristan Hubsch, 
Raghu Mahajan, Jacob McNamara and Timo Weigand 
for useful discussions.
This work was supported by the ICTS-Infosys Madhava 
Chair Professorship
and the Department of Atomic Energy, Government of India, under project no. RTI4001.
I would also like to acknowledge the hospitality of the Department of Physics,
Imperial College, London and the Department of Mathematics,
King's College, London for hospitality during the course of this work.

\end{document}